\documentclass[twocolumn]{aastex63}
\usepackage{amsmath}
\usepackage{enumerate}

\begin{document}

\title{Examining the Relationship Between Convective Core Overshoot and Stellar Properties Using Asteroseismology}

\author[0000-0002-7541-9346]{Lucas S. Viani}
\affiliation{Department of Astronomy, Yale University, New Haven, CT, 06520, USA}

\author[0000-0002-6163-3472]{Sarbani Basu} 
\affiliation{Department of Astronomy, Yale University, New Haven, CT, 06520, USA}

\email{lucas.viani@yale.edu}

\begin{abstract}
Core overshoot is a large source of uncertainty in constructing stellar models. Whether the amount of overshoot is constant or mass dependent is not completely known, even though models sometimes assume a mass-based trend. In this work we use asteroseismic data from stars observed by \textit{Kepler} to investigate the relationship between various stellar properties and the amount of overshoot needed to properly model a given star. We find a strong positive trend between stellar mass and overshoot amount for stars between 1.1 and 1.5 $M_\odot$, with a slope of 0.89. Additionally, we investigate how inferred stellar properties change as a function of overshoot. Our model grids show that the inferred stellar mass and radius can vary by as much as 14\% and 6\% respectively, depending on the extent of overshoot. This mass spread results in a commensurate spread in the ages.
\end{abstract}


\section{Introduction}

One of the largest sources of uncertainty in stellar models is the implementation of convection. Due to the short convective timescales compared to the timescales over which the stars are being evolved, several approximations must be made. In many modeling codes, these approximations result in the convective process being controlled by two free parameters: the mixing-length parameter \citep{BohmVitense1958} and the parameter that determines the extent of overshoot, particularly from convective cores. These values set the efficiency of convection and how far past the convective boundary a bubble is carried due to its velocity and momentum. While the use of these free parameters make modeling convection much more convenient, the values that should be used when constructing stellar models are uncertain. Without knowledge of the value of the overshoot parameter or mixing-length parameter one must either make an educated guess, which may then produce inaccurate results, or create many more sets of models to account for the extra free dimensions. For a more detailed discussion regarding the difficulties that arise in stellar modeling due to the handling of convection see \cite{Kippenhahn2012}, \cite{Salaris2017}, \cite{BasuBook2017}, and references therein.

In \cite{Viani2018} we examined the mixing-length parameter using asteroseismology. In this work we use asteroseismology to explore the relationship between various stellar properties and the overshoot value needed to properly model a set of stars. Understanding the relationships between stellar properties and the required overshoot amount will greatly reduce the ambiguity and uncertainty on the value of the overshoot parameter and will reduce the uncertainty in inferred stellar parameters caused by uncertainties in overshoot. Additionally, we explore the impact that different overshoot amounts have on inferred stellar parameters. By assuming various fixed overshoot values for the stars we can track the sensitivity of derived stellar properties to overshoot. 

Convective overshoot refers to the phenomenon where a rising convective bubble, moving towards the edge of the convective zone, is carried past the convective boundary due to the bubble's momentum. While the region outside the convective zone is stable against convection, since the bubble has mass and velocity approaching the boundary it will ``overshoot'' and move some distance beyond the convective zone. In models, the distance the bubble can overshoot is conventionally modeled as $\alpha_\mathrm{ov} H_\mathrm{P}$, where $H_\mathrm{P}$ is the pressure scale height and $\alpha_\mathrm{ov}$ is the overshoot parameter. While overshooting can happen in both envelope and core convective zones, for the purposes of this work we are only concerned with convective core overshoot in stars that have a convective core.
 
The inclusion of overshoot in stellar models is important for many reasons. First of all, overshoot is expected because the criteria for determining the convective boundary give a condition on acceleration and not velocity, and hence the convective eddies are expected to continue moving into the stable layer. Second, core overshoot is going to impact the star's main-sequence lifetime and alter the main-sequence turnoff location in isochrones. Since the overshooting region causes mixing beyond the convective core, its inclusion effectively increases the size of the core thereby increasing the amount of hydrogen available for fusion and consequently increasing the main-sequence lifetime of the star or model. Stellar tracks and isochrones with and without overshoot have substantial differences at turn-off, on the subgiant branch, and also at the base of the red giant branch \citep[see, e.g.,][]{Prather1974, Maeder1976, Maeder1981, Chin1991, Stothers1991, Woo2001, Yi2004, Demarque2004, Pietrinferni2004, Salaris2017}. Many studies have suggested that the inclusion of overshoot in the models may be needed in order to properly fit many open cluster color magnitude diagrams \citep[CMDs; e.g.,][and others]{Maeder1981, Carraro1993, Daniel1994, Demarque1994, Kozhurina1997, Nordstroem1997, Barmina2002, Woo2003, Pietrinferni2004}. This is due to many clusters having a characteristic ``hook'' feature near turn-off as well as a luminosity gap which are both better reproduced with overshoot. Attempts to tightly constrain the overshoot amount can be difficult due to uncertainties in helium abundance and metallicity \citep[see, for example,][]{VandenBerg2007}. 

While the need to consider the inclusion of overshoot is apparent, what is less clear is what the value of the overshoot parameter should be. Fits to cluster CMDs suggest that the overshoot should be around 0.2$H_P$ \citep[e.g.,][etc.]{Maeder1989,Demarque1994,Kozhurina1997}. Typically, this value is used, however, the overshoot amount needed varies on a star by star (or cluster by cluster) basis. The current convention is to implement an overshoot scheme where for massive stars the overshoot amount is around 0.2$H_P$ and for lower mass stars the overshoot parameter increases with stellar mass \citep{Demarque2004,Pietrinferni2004,Bressan2012,Hidalgo2018}. Since these less massive stars have smaller convective cores, they are expected to have smaller overshoot. 

There are earlier, non-asteroseismic, empirical studies of the link between mass and overshoot. A series of papers by \cite{Claret2016,Claret2017,Claret2018,Claret2019}, used a set of 50 eclipsing binaries to determine the overshoot amounts in 100 stars. These studies found a positive trend between stellar mass and overshoot for stars with masses up to about $2 M_{\odot}$. However, several studies have questioned the reliability of using eclipsing binaries to constrain the amount of overshoot \citep{Constantino2018,Valle2016,Valle2017,Valle2018}, thus cautioning about interpreting the apparent mass trend. A study by \cite{Stancliffe2015} modeled 12 eclipsing binary systems of masses between 1.3 and 6.2 $M_\odot$ and found no trend between overshoot and mass. 

Asteroseismology allows us to probe the internal structure of stars, and hence should be able to provide more reliable results. One reason that the results from eclipsing binaries must be interpreted with caution is due to uncertainties in the metallicity and elemental abundances and distributions. While this can still be an issue with asteroseismology, an advantage is that there is much more available observational seismic data. With eclipsing binaries there is typically only knowledge of a few parameters, i.e., $T_\mathrm{eff}$, metallicity, mass, and radius, while with asteroseismology each frequency is a data point that can be matched to the models.

\cite{Aguirre2011} performed a theoretical study showing that seismic data can be used to determine the existence of a convective core and differentiate the size of the mixed region. An additional advantage is that seismology can be applied to single stars. Many other studies have made use of asteroseismology to probe the extent of the convective core and overshoot region in stars, such as, \cite{Deheuvels2010}, \cite{Degroote2010}, \cite{Deheuvels2011}, \cite{Neiner2012}, \cite{Montalban2013}, \cite{Aguirre2013}, \cite{Guenther2014}, \cite{Aerts2015}, \cite{Moravveji2015}, \cite{Deheuvels2016}, \cite{Angelou2020}, and many others.

Using asteroseismic data, \cite{Deheuvels2016} determined the overshoot amount in a set of \textit{Kepler} stars. Their results show a positive trend between mass and overshoot. However, not all studies have found such clear trends. \cite{Angelou2020} use \textit{Kepler}, \textit{CoRoT}, and radial velocity data to determine overshoot values. While they do see a range in overshoot for their set of stars, they do not see a strong trend between a star's overshoot and mass. Thus, clearly, additional investigation into the relationship between stellar mass and overshoot is needed.

In this work we will further investigate the possible link between mass and overshoot, as well as investigate possible trends between overshoot and other stellar parameters. We will do this by modeling a set of stars from the \textit{Kepler} Asteroseismic LEGACY Sample \citep{Lund2017,Aguirre2017}. Through detailed modeling we will determine the overshoot amount needed to best reproduce each star, making use of the sample's asteroseismic data.

Additionally, we will use the models to investigate how inferred stellar properties change depending on the overshoot value. In other words, if just one overshoot value is assumed, and only that subgrid is used to determine the stellar properties, how much do the inferred stellar properties change? This is important in understanding the impact that the overshoot parameter has on derived stellar properties and to understand which stellar properties are the most sensitive to overshoot. This will also allow us to examine the potential importance of including multiple overshoot values in modeling studies, instead of the common practice of using only one single value.

The rest of the paper is organized as follows. Section~\ref{sec:data} describes the sample of stars, the construction of the stellar models, and the analysis techniques used to determine stellar properties from the model grids. Section~\ref{sec:results} shows the relationship between stellar properties and overshoot amount, investigates how much the inferred stellar properties differ depending on overshoot, examines using different overshoot physics, and explores using frequency separation ratios as a predictor of overshoot. Section~\ref{sec:conclusion} contains concluding remarks.

\section{Data, Models, and Analysis}
\label{sec:data}
\subsection{Sample of Stars}
\begin{figure}
\epsscale{1.0}
\plotone{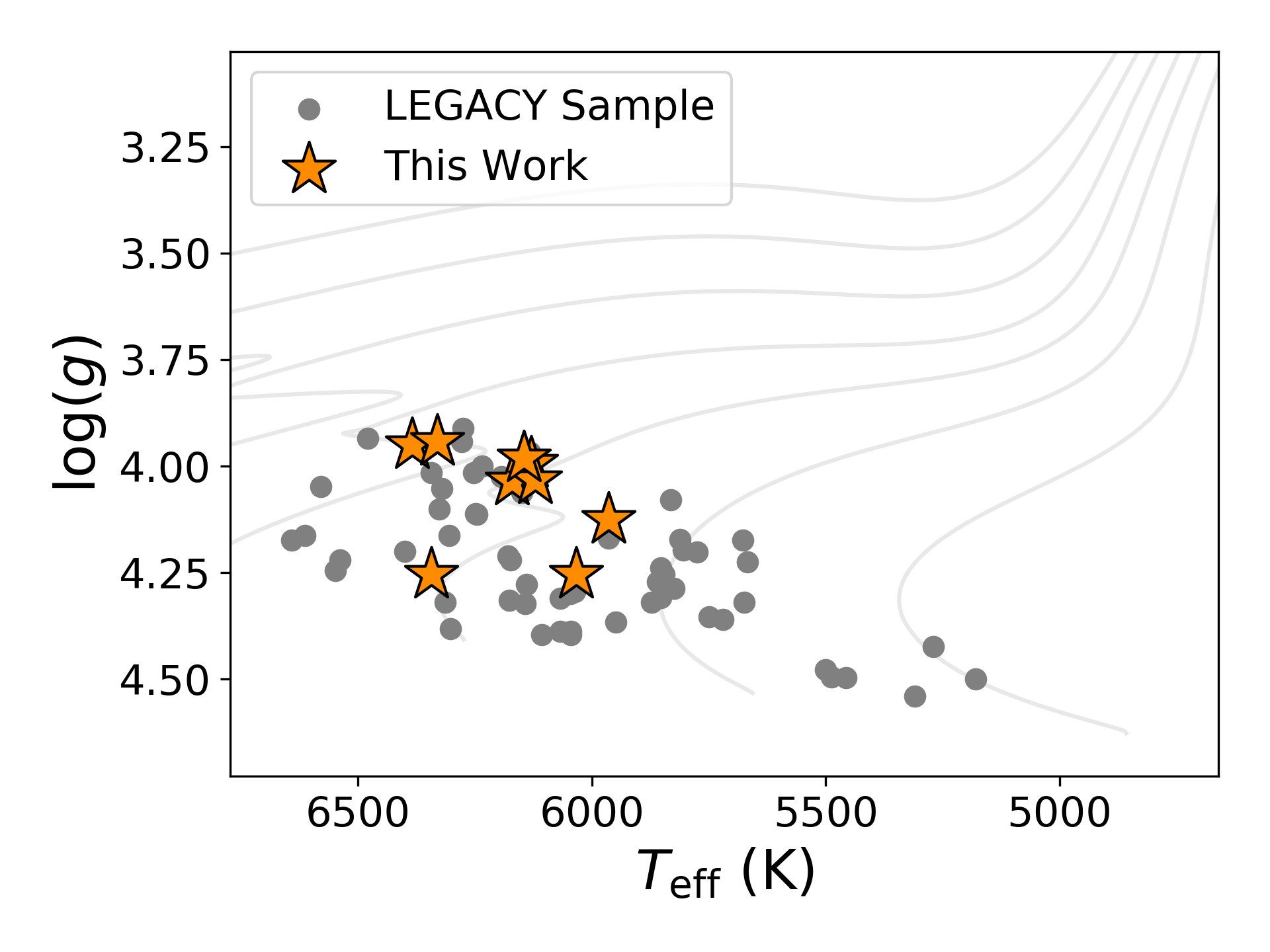}
\caption{The stars in this work (orange) compared to the full LEGACY sample (gray).}
\label{fig:HR}
\end{figure}

We use a subset of stars from the \textit{Kepler} Asteroseismic LEGACY Sample from \cite{Lund2017} and \cite{Aguirre2017}. The specific stars included in this work were randomly selected from the higher mass subset of the sample ($>1.2$ $\mathrm{M}_\odot$) to ensure that they would have a convective core. A few lower mass stars were also included to cover a wider range of parameter space. The full list of the stars examined in this work can be seen in Table~\ref{Table:Stellar_Properties}. Figure~\ref{fig:HR} shows our sample of stars plotted on a Kiel diagram.

\subsection{Constructing the Models}
To determine the amount of overshoot needed to best model each of the stars, grids of stellar models were generated using YREC \citep{Demarque2008}. The models were constructed using the OPAL equation of state \citep{Rogers2002} and OPAL opacities \citep{Iglesias1996} with low temperature opacities from \cite{Ferguson2005}. The \cite{Adelberger1998} nuclear reaction rates were used, except for the $^{14}N$($p$,$\gamma$)$^{15}O$ reaction, where the rates of \cite{Formicola2004} were implemented. The models use the \cite{GS1998} heavy element distribution. Additionally, Eddington gray atmospheres were used. These models do not include the diffusion and gravitational settling of helium and heavy elements. 

An independent grid was constructed for each star. To get an estimate of the parameter space each grid needed to cover, Table 4 from \cite{Aguirre2017} was used. Here the LEGACY stars were modeled using a variety of pipelines with each pipeline returning their best-fit stellar properties. We first estimated the mass of the star, $M_\mathrm{est.}$, by taking the mean of the various mass values from \cite{Aguirre2017}. The mass range of our grid then span from $M_\mathrm{est.} \pm 2 \sigma_{M}$, where $\sigma_{M}$ is the mass uncertainty, taken to be the largest uncertainty for that star in \cite{Aguirre2017} across all the pipelines. The step size, $\Delta M$ was 0.02 $M_\odot$. This typically resulted in about 20 different masses per star. A similar approach was used to estimate the star's metallicity. The estimated $\mathrm{[Fe/H]}$ was determined by averaging the values of the various pipelines in \cite{Aguirre2017}. Our grid used 7 different metallicities, ranging from $\mathrm{[Fe/H]_{est.}} \pm 2\sigma_\mathrm{[Fe/H]}$. The mixing-length parameter, $\alpha_{\mathrm{MLT}}$ ranged from 1.2 to 2.7 in steps of 0.25. The initial helium abundance, $Y_0$, ranged from 0.248 to 0.356 in steps of 0.018. A range of core-overshoot values were also used in each grid. We use the step overshoot implementation for our models. The number of different overshoot values used depended on the star and how the posterior likelihood distribution was shaped, but typically about 8 different values are used per star, in steps of 0.1. 

For our default grid, discussed in the first half of this work, the overshoot prescription is such that the temperature gradient of the overshooting region is adiabatic. The results of using different overshoot physics are discussed later in Section~\ref{sec:model_physics}.

\subsection{Input vs. Effective Overshoot}
It is important that we differentiate between the input overshoot for a model and the actual value of the overshoot, which we will refer to as the ``effective'' overshoot. The input overshoot value is what is entered into the modeling code as the free parameter, while the effective overshoot measures the actual resulting overshoot amount in the model. For stars in the mass range of our sample, these values are often significantly different. The reason that these two values can differ is due to safeguards in modeling codes that prevent unrealistically large  overshoot regions. While in simple terms the overshoot radius is defined as,
\begin{equation}
\label{eq:R_ov}
R_\mathrm{ov} = \alpha_\mathrm{ov} H_P
\end{equation}
this can, in some cases, result in overshooting regions that are larger than the convective core. This is especially the case in lower mass stars. As $r \rightarrow 0$ then $H_P \rightarrow \infty$, which means that since lower mass stars have smaller convective cores, the value of $\alpha_{\mathrm{ov}}$ must be reduced or the overshoot region will become unrealistically large \citep{Roxburgh1992,Woo2001}. To avoid this issue, the actual overshoot radius is defined by YREC as,
\begin{equation}
\label{eq:R_eff}
R_\mathrm{ov} = \frac{\alpha_\mathrm{ov}}{(1/H_P)+(1/R_\mathrm{cz})}
\end{equation}
where $R_\mathrm{cz}$ is the radius of the convective core. If the convective core of the star is large, than Eq.~\ref{eq:R_eff} simplifies to Eq.~\ref{eq:R_ov}. Since YREC uses Eq.~\ref{eq:R_eff} to determine the size of the overshoot region, an input value of $\alpha_\mathrm{ov}=0.2$ does not necessarily translate to an overshoot radius of 0.2$H_P$. Thus, while the input overshoot value is 0.2, the effective overshoot value will be lower, depending on the value of $R_\mathrm{cz}$. Similar safeguards are also present in other modeling codes, for example MESA \citep{Paxton2018} and GARSTEC \citep{Weiss2008}.

\begin{figure}
\epsscale{1.0}
\plotone{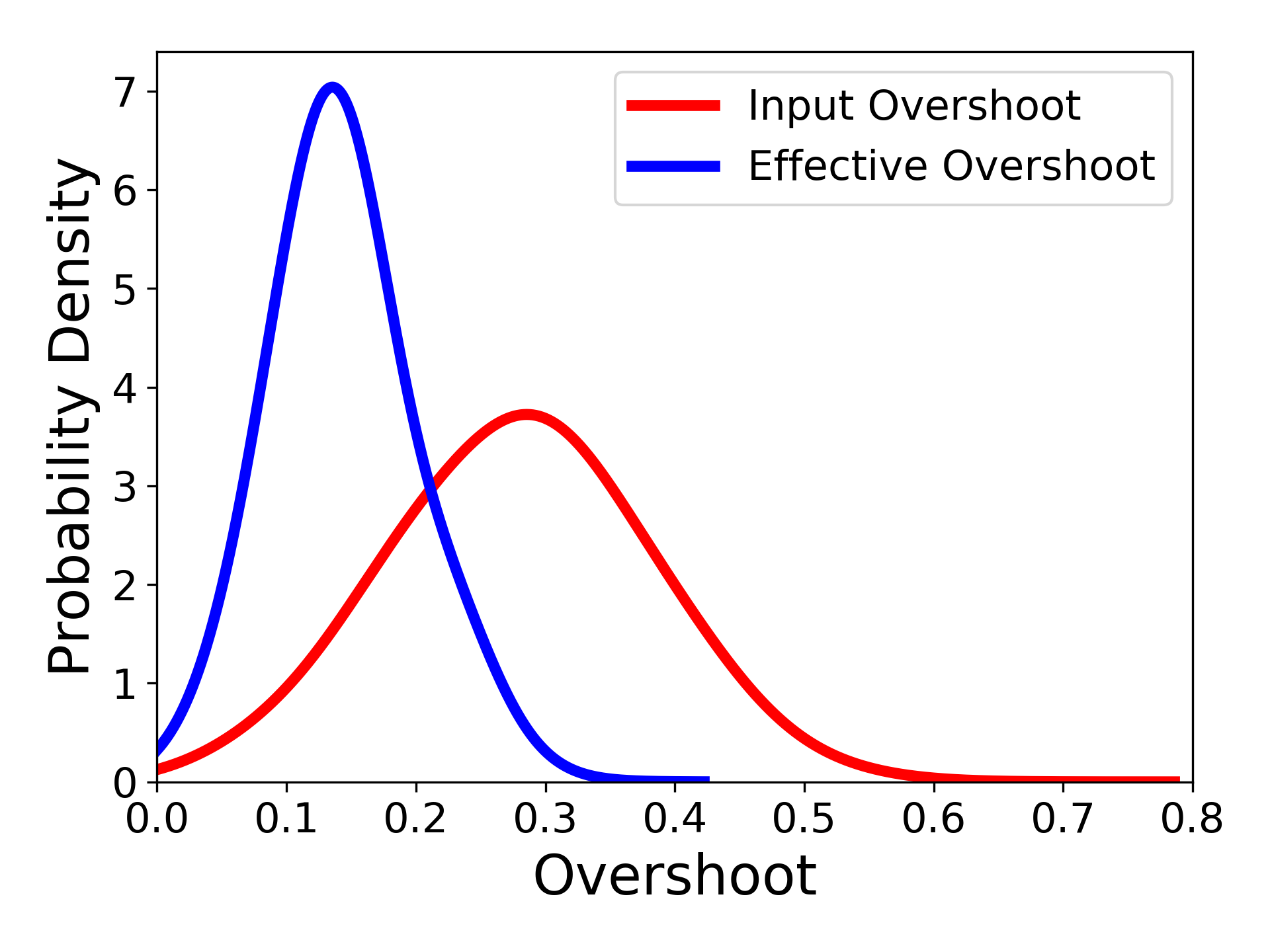}
\caption{The probability density for input overshoot (red) and effective overshoot (blue) for KIC 8228742. Overshoot is in units of $H_P$.}
\label{fig:input_vs_effective_PDF}
\end{figure}

Figure~\ref{fig:input_vs_effective_PDF} shows the probability density function (PDF) for KIC 8228742 for both measurements of overshoot. Clearly, the PDF of the effective overshoot peaks at a significantly lower value compared to the input overshoot. In this set of model grids, input overshoot values up to 0.8 were used. However, in the effective overshoot space, this translates to an overshoot of around 0.4. The potentially large differences between the input and effective overshoot values are also illustrated in \cite{Angelou2020}. While the input overshoot value is important, as that is what can be controlled when constructing the stellar models, the effective overshoot value is the actual extent of the overshooting region. The input overshoot is simply what is supplied as a free parameter to the modeling code, while the true size of the overshooting region in the model is given by the effective overshoot value. For the rest of this work, when we refer to overshoot, we mean the effective overshoot value, unless otherwise specified.    

\subsection{Determining Stellar Properties from Model Grids}
For each grid, we calculate the $\ell=0,$ $1,$ and $2$ frequencies for every model that is within $\pm 4 \sigma$ of the estimated $T_\mathrm{eff}$ and $\log g$. We then calculate the $r_{01}$ and $r_{02}$ frequency ratios, which will be used in the model likelihood calculations. These frequency ratios are given by Eqs.~\ref{eq:r01} and \ref{eq:r02} \citep[see, e.g.,][]{BasuBook2017}
\begin{equation}
\label{eq:r01}
r_{01}(n)= \frac{\nu_{n,0} - \frac{1}{2}(\nu_{n-1,1} + \nu_{n,1})}{\nu_{n,1} - \nu_{n-1,1}}
\end{equation} 
and
\begin{equation}
\label{eq:r02}
r_{02}(n)= \frac{\nu_{n+1,0} - \nu_{n,2}}{\nu_{n,1} - \nu_{n-1,1}}.
\end{equation}
These ratios are especially useful because they avoid the need for surface-term corrections that would otherwise have to be applied before comparing model and observed frequencies \citep{Roxburgh2003}. We have ignored the correlation between the separation ratios; our preliminary investigations showed that the $\chi^2$ is relatively insensitive. Assuming uncorrelated errors speeds up the calculations many fold, which is important since we test millions of models.

A chi-squared value for each model is then determined, following Eq.~\ref{eq:Chi}, based on the model's frequency ratios, $T_\mathrm{eff}$, and $\mathrm{[Fe/H]}$. The $\chi^2$ value is calculated as,
\begin{equation}
\label{eq:Chi}
\chi^2_\mathrm{total}= \chi^2_{r_{01}} + \chi^2_{r_{02}} + \chi^2_{\mathrm{[Fe/H]}} + \chi^2_{T_{\mathrm{eff}}}
\end{equation}
where   
\begin{equation}
\label{eq:Chi_01}
\chi^2_{r_{01}}=\left(\sum \frac{(r_{01,\mathrm{obs}} - r_{01,\mathrm{model}})^2}{\sigma^2_{r_{01,\mathrm{obs}}}}\right) \left( \frac{1}{n_{\mathrm{modes}}-1} \right),
\end{equation}
\begin{equation}
\label{eq:Chi_02}
\chi^2_{r_{02}}=\left(\sum \frac{(r_{02,\mathrm{obs}} - r_{02,\mathrm{model}})^2}{\sigma^2_{r_{02,\mathrm{obs}}}}\right)  \left(\frac{1}{n_{\mathrm{modes}}-1} \right),
\end{equation}
\begin{equation}
\label{eq:Chi_FeH}
\chi^2_\mathrm{[Fe/H]} = \frac{\left(\mathrm{[Fe/H]}_\mathrm{est.} - \mathrm{[Fe/H]}_\mathrm{model}\right)^2}{\sigma^2_\mathrm{[Fe/H]}},
\end{equation}
and
\begin{equation}
\label{eq:Chi_Teff}
\chi^2_{T_\mathrm{eff}} = \frac{\left(T_\mathrm{eff,est.} - T_\mathrm{eff,model}\right)^2}{\sigma^2_{T_\mathrm{eff}}}.
\end{equation}

\begin{figure}
\epsscale{1.0}
\plotone{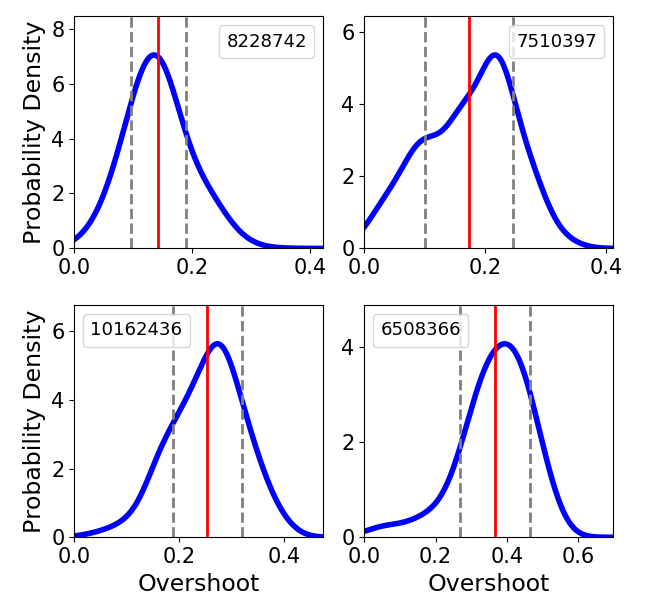}
\caption{The probability density function for effective overshoot (in units of $H_P$) for four of the stars in the sample is shown in blue. The red line shows the likelihood weighted average value (Eq.~\ref{eq:Like_weighted}) and the gray lines show the uncertainty (Eq.~\ref{eq:likelihoods_uncert}). The number label indicates the star's KIC number.}.
\label{fig:KDE_example}
\end{figure}

The total likelihood for each model is then given by
\begin{equation}
\label{eq:Likes}
\mathcal{L}_\mathrm{total} \propto e^{-\chi^2_\mathrm{total}  / 2}.
\end{equation}
To determine stellar properties we then use a likelihood weighted average. For a given stellar parameter, the likelihoods are converted into a probability density by marginalizing over the other properties and dividing each bin by the number of stars in the bin. Using this probability density information, $\mathcal{P}$, the weighted average is then calculated for a wide variety of stellar properties following Eq.~\ref{eq:Like_weighted}, using overshoot as an example, 
\begin{equation}
\label{eq:Like_weighted}
\left\langle \alpha_\mathrm{ov}\right\rangle =\frac{\sum \alpha_{\mathrm{ov},i} \mathcal{P}_i}{\sum \mathcal{P}_i}.
\end{equation}
The uncertainty in each property is determined by,
\begin{equation}
\sigma^2 = \frac{\sum (\alpha_{\mathrm{ov},i}-\langle \alpha_\mathrm{ov} \rangle)^2 \mathcal{P}_i }{\sum \mathcal{P}_i}.
\label{eq:likelihoods_uncert}
\end{equation}
In this manner the stellar properties were determined from each star's model grids. A visual example of the likelihood weighted average and the uncertainties can be seen in Figure~\ref{fig:KDE_example}, where we show the probability density functions for overshoot for four of the stars. Plotted over the PDFs are the calculated likelihood weighted averages and uncertainties. As can be seen, the PDF peak and the likelihood weighted average agree very well. The small shift between the two values in the upper right panel is due to the slightly asymmetric PDF for that star.
 
\section{Results and Discussion}
\label{sec:results}
\subsection{Relationships Between Overshoot and Stellar Properties}
\begin{figure*}
\epsscale{1.0}
\plotone{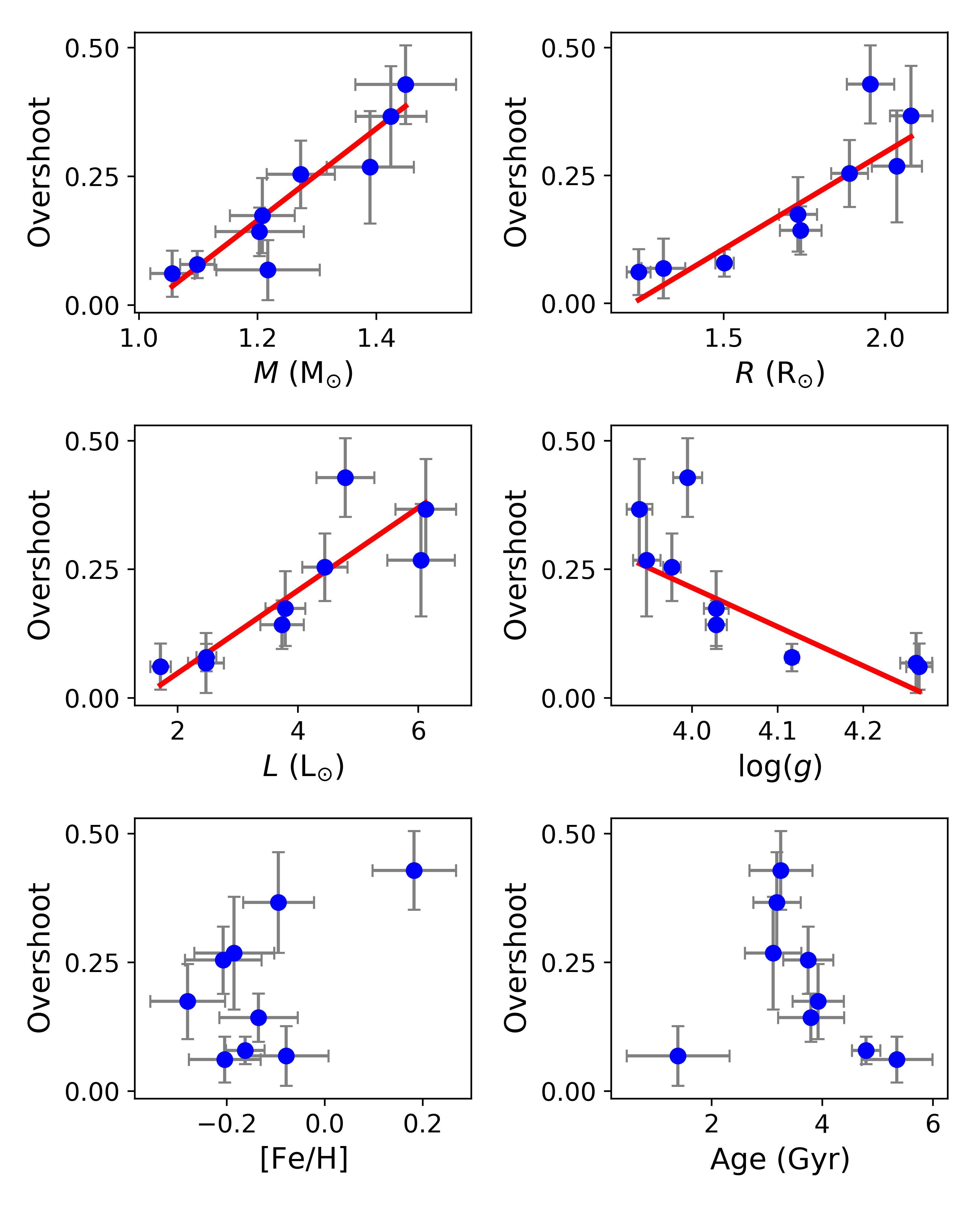}
\caption{The effective overshoot value (in units of $H_P$) plotted as a function of a variety of stellar properties for the stars in the study. The red line is a weighted linear best fit. A version of this figure created using the input overshoot can be seen in Fig.~\ref{fig:input_overshoot_multiplot} in the Appendix.}
\label{fig:overshoot_multiplot}
\end{figure*}

With the inferred properties for each star determined from the grids of models, we then investigate the relationship between overshoot and various stellar properties. Figure~\ref{fig:overshoot_multiplot} shows overshoot plotted as a function of a few properties.
From Fig.~\ref{fig:overshoot_multiplot} there is a clear positive trend between the star's mass and the overshoot amount. The slope of the weighted line of best fit is 0.89. There are also trends between overshoot and radius, luminosity, and $\log g$, however these trends likely arise due to their correlation with stellar mass. We do not observe a trend between overshoot and mixing length, $\mathrm{[Fe/H]}$, age, $T_\mathrm{eff}$, or $Y_0$. It may be that the sample size is too small to detect such trends between overshoot and these other stellar properties. The inferred properties for the set of stars are listed in Table~\ref{Table:Stellar_Properties}.

\begin{table*}[]
\centering
\caption{The derived values for a variety of stellar properties, including effective overshoot, for the stars in the sample. The table is ordered by the inferred stellar mass.}
\begin{tabular}{ccccccccc}
\hline
\hline
KIC & Mass ($M_\odot$) & Age (Gyr) & $T_\mathrm{eff}$ (K) & $\log g$ & $\mathrm{[Fe/H]}$ & $Y_0$ & Mixing Length ($H_P$) & Overshoot ($H_P$) \\ \hline
6116048 &	1.06$\pm$0.04 & 5.35$\pm$0.64 & 5972$\pm$81 & 4.265$\pm$0.015 & -0.20$\pm$0.07 & 0.260$\pm$0.016 & 1.45$\pm$0.15 & 0.06$\pm$0.04 \\
12258514 & 	1.10$\pm$0.03 & 4.79$\pm$0.26 & 5936$\pm$65 & 4.117$\pm$0.007 & -0.16$\pm$0.04 & 0.276$\pm$0.017 & 1.26$\pm$0.11 & 0.08$\pm$0.03 \\
8228742 & 	1.20$\pm$0.07 & 3.79$\pm$0.60 & 6096$\pm$79 & 4.028$\pm$0.012 & -0.14$\pm$0.08 & 0.299$\pm$0.031 & 1.39$\pm$0.11 & 0.14$\pm$0.05 \\
7510397 & 	1.21$\pm$0.05 & 3.92$\pm$0.46 & 6136$\pm$78 & 4.028$\pm$0.015 & -0.28$\pm$0.08 & 0.279$\pm$0.026 & 1.36$\pm$0.13 & 0.17$\pm$0.07 \\
8179536 & 	1.22$\pm$0.09 & 1.39$\pm$0.93 & 6313$\pm$79 & 4.262$\pm$0.019 & -0.08$\pm$0.09 & 0.310$\pm$0.032 & 1.36$\pm$0.18 & 0.07$\pm$0.06 \\
10162436 & 	1.27$\pm$0.06 & 3.74$\pm$0.45 & 6109$\pm$79 & 3.976$\pm$0.010 & -0.21$\pm$0.08 & 0.271$\pm$0.023 & 1.39$\pm$0.14 & 0.25$\pm$0.07 \\
3456181 & 	1.39$\pm$0.07 & 3.11$\pm$0.51 & 6353$\pm$79 & 3.947$\pm$0.016 & -0.18$\pm$0.08 & 0.277$\pm$0.026 & 1.53$\pm$0.23 & 0.27$\pm$0.11 \\
6508366 & 	1.42$\pm$0.06 & 3.18$\pm$0.43 & 6297$\pm$78 & 3.938$\pm$0.015 & -0.09$\pm$0.07 & 0.270$\pm$0.024 & 1.54$\pm$0.21 & 0.37$\pm$0.10 \\
5773345 & 	1.45$\pm$0.08 & 3.25$\pm$0.57 & 6094$\pm$86 & 3.995$\pm$0.017 & \phantom{-}0.18$\pm$0.09 & 0.289$\pm$0.030 & 1.37$\pm$0.15 & 0.43$\pm$0.08 \\ \hline
\end{tabular}
\label{Table:Stellar_Properties}
\end{table*}

\begin{figure}
\epsscale{1.0}
\plotone{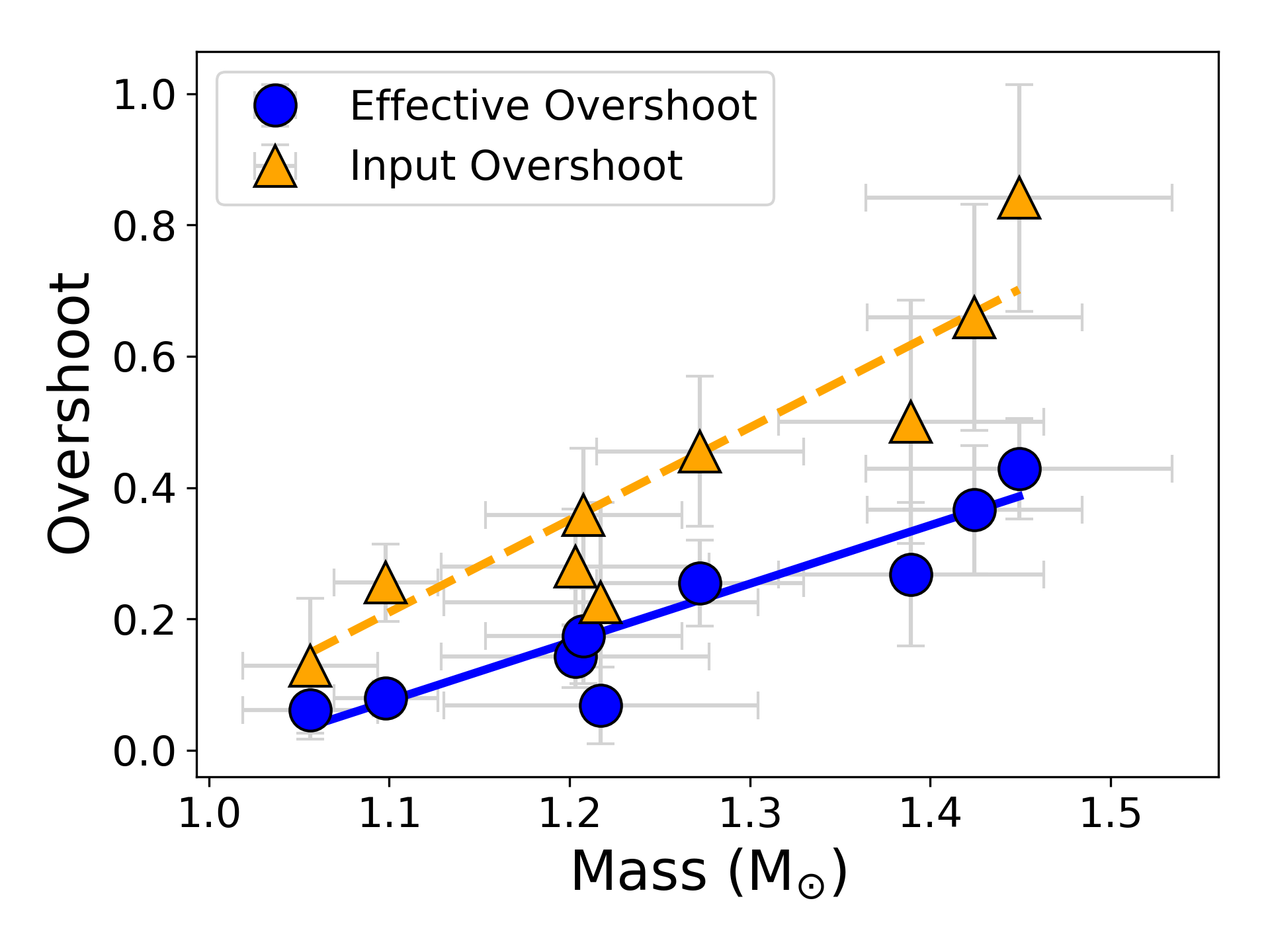}
\caption{The input overshoot (orange points) and effective overshoot (blue points) as a function of mass for the stars in the study. Overshoot is in units of $H_P$.}
\label{fig:input_vs_effective}
\end{figure}

It is also of interest to examine this trend between mass and overshoot in regards to the input overshoot value, instead of the effective overshoot amount. In Figure~\ref{fig:input_vs_effective} we plot both sets of overshoot values as a function of mass. As seen in Fig.~\ref{fig:input_vs_effective} the positive trend with mass is present in both cases. The slope of the trend is less for the effective overshoot case. As expected, the effective overshoot values are less than the input overshoot values, with the difference being greater at higher masses. The full set of panels from Fig.~\ref{fig:overshoot_multiplot} are remade but using the input overshoot instead of the effective overshoot in Fig.~\ref{fig:input_overshoot_multiplot} in the Appendix.

\subsection{How Inferred Stellar Properties Change with the Amount of Overshoot}
It is also important to examine how the inferred stellar properties change if only one single overshoot value is used in the modeling process. This is crucial, because many modeling studies often only assume a single, fixed, overshoot value. If we see that inferred stellar properties are very sensitive to the overshoot amount, then the use of multiple overshoot values in modeling studies will need to become more commonplace. 
 
\begin{figure*}
\epsscale{1.0}
\plotone{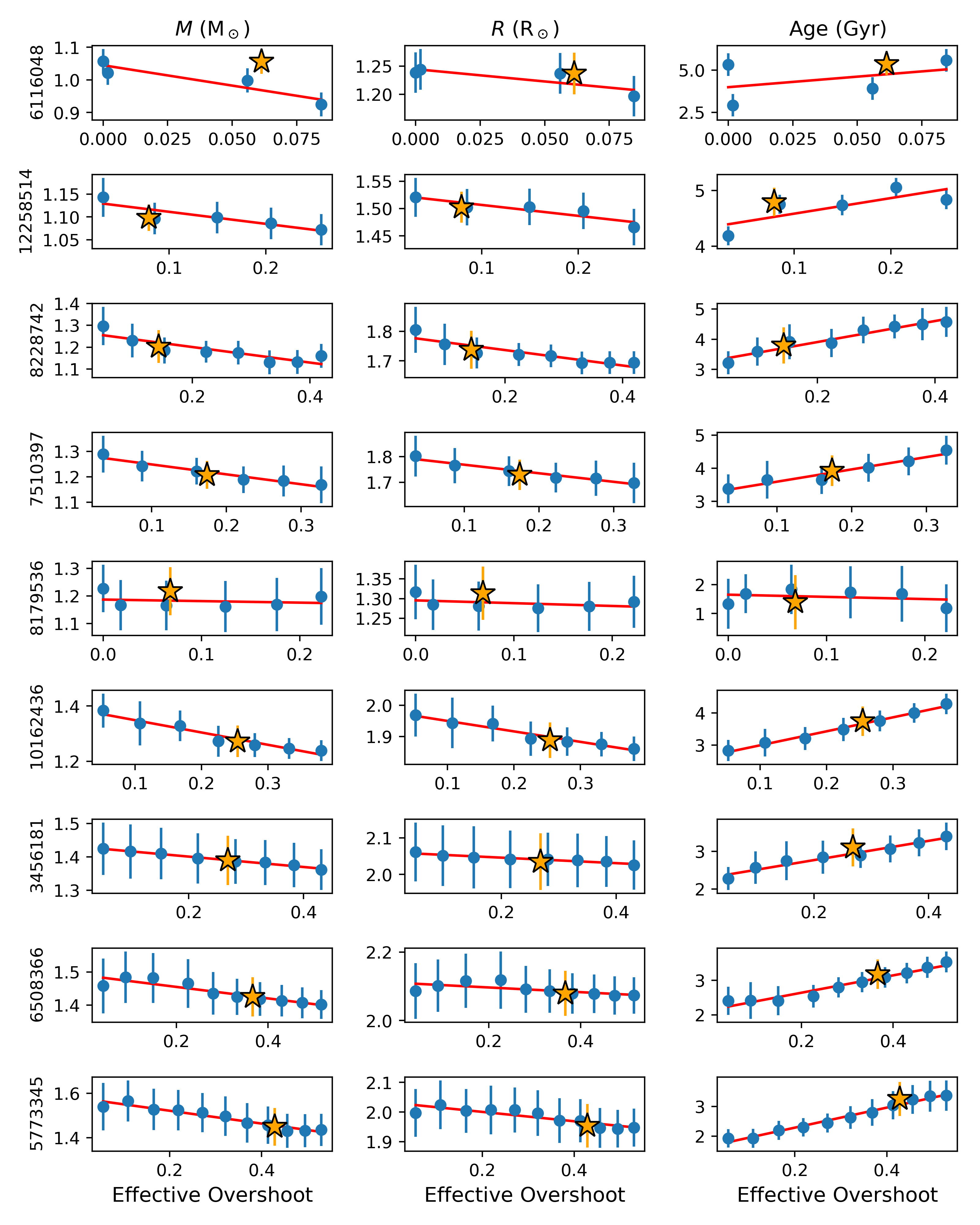}
\caption{The inferred values of various stellar properties for each of the overshoot subgrids for each star. Column 1 shows the mass, Column 2 the radius, and Column 3 the age. The various rows, labeled by KIC number, are the different stars in the study. The orange star symbol in each panel is the inferred value for all subgrids combined. The red line is a line of best fit to help guide the eye. Overshoot is in units of $H_P$.}
\label{fig:individual_grids}
\end{figure*}

Using just one overshoot subgrid at a time, we calculate the inferred stellar properties. The results are plotted in Figure~\ref{fig:individual_grids}. From Fig.~\ref{fig:individual_grids} we see that the derived values of mass, radius, and age can change greatly depending on the grid's overshoot value. The range in inferred mass, radius, and age values for the different stars can be seen in Table~\ref{Table:Inferred_Spread}. Here we see that the difference in inferred mass can be as high as 14\%, the difference in inferred radius as large as 6\%, and the difference in inferred age as large as 50\%, depending on the overshoot value of the models. The inferred age is by far the most affected property, which makes sense due to the fact that overshoot is effectively altering the amount of fuel available to the core. These large differences in derived stellar properties highlight the key importance of selecting the proper value of overshoot for one's model grids.  

\begin{table*}[]
\centering
\caption{The range in inferred mass, radius, and age based on using one overshoot subgrid at a time. This table provides the numerical spread for the data presented in Fig.~\ref{fig:individual_grids}. For each property, the first column gives the inferred value from the ensemble of all overshoot grids, the second column is the range in inferred values using only one overshoot subgrid at a time, and the third column gives the range percent: $100 \times (\mathrm{highest} - \mathrm{lowest}) / (\mathrm{inferred \: ensemble \: value})$.}
\begin{tabular}{cccccccccc}
\hline
\hline
KIC      & Mass ($M_\odot$) & $M_\mathrm{Range}$ & $M_\mathrm{Range} (\%)$ & Radius ($R_\odot$)    & $R_\mathrm{Range}$   & $R_\mathrm{Range} (\%)$ & Age (Gyr) & $\mathrm{Age_{Range}}$ & $\mathrm{Age_{Range}} (\%)$ \\ \hline
6116048  & 1.06 & 0.92--1.06  & 12.6   & 1.24 & 1.20--1.24 & 3.9     & 5.35 & 2.91--5.57 & 49.7  \\ 
12258514 & 1.10 & 1.07--1.14  & 6.4    & 1.50 & 1.47--1.52 & 3.6     & 4.79 & 4.18--5.05 & 18.1  \\
8228742  & 1.20 & 1.13--1.30  & 13.8    & 1.74 & 1.69--1.81 & 6.5     & 3.79 & 3.22--4.57 & 35.6 \\
7510397  & 1.21 & 1.17--1.29  & 10.0    & 1.73 & 1.70--1.80 & 6.0     & 3.92 & 3.38--4.54 & 29.7 \\
8179536  & 1.22 & 1.16--1.23  & 5.4    & 1.31 & 1.28--1.32 & 3.1     & 1.39 & 1.18--1.83 & 46.7 \\
10162436 & 1.27 & 1.24--1.38  & 11.4    & 1.89 & 1.86--1.97 & 5.7     & 3.74 & 2.84--4.28 & 38.5 \\
3456181  & 1.39 & 1.36--1.42  & 4.5    & 2.04 & 2.03--2.06 & 1.8     & 3.11 & 2.28--3.40 & 36.2  \\
6508366  & 1.42 & 1.40--1.48  & 5.8    & 2.08 & 2.07--2.12 & 2.2     & 3.18 & 2.41--3.54 & 35.5 \\
5773345  & 1.45 & 1.43--1.57  & 9.3    & 1.95 & 1.94--2.02 & 4.1     & 3.25 & 1.93--3.36 & 44.2  \\ \hline
\end{tabular}
\label{Table:Inferred_Spread}
\end{table*}

Since the inferred stellar properties can change so drastically depending on the assumed overshoot value, future modeling work needs to be careful to include a wide range of overshoot amounts. Current practice often assumes one overshoot value, or in some cases one set of models without overshoot and one set with a single fixed overshoot value. For example, many of the pipelines used to model stars in the \cite{Aguirre2017} LEGACY paper use only one set of input overshoot values in the models, such as AIMS \citep{Reese2016,Rendle2019}, C2kSMO \citep{Lebreton2014}, and the YMCM method described in \cite{Aguirre2015} that used the same YREC code that we use here. Additionally, \cite{Viani2018} assumed either no overshoot or an input value of 0.2$H_P$. Clearly, from Table~\ref{Table:Inferred_Spread}, assuming only one overshoot value is dangerous, as the inferred stellar properties can differ greatly depending on the assumed overshoot amount. Therefore, to obtain the most reliable results, it is important to treat overshoot as a free parameter and include a wide range of overshoot values.  

\begin{figure}
\epsscale{1.0}
\plotone{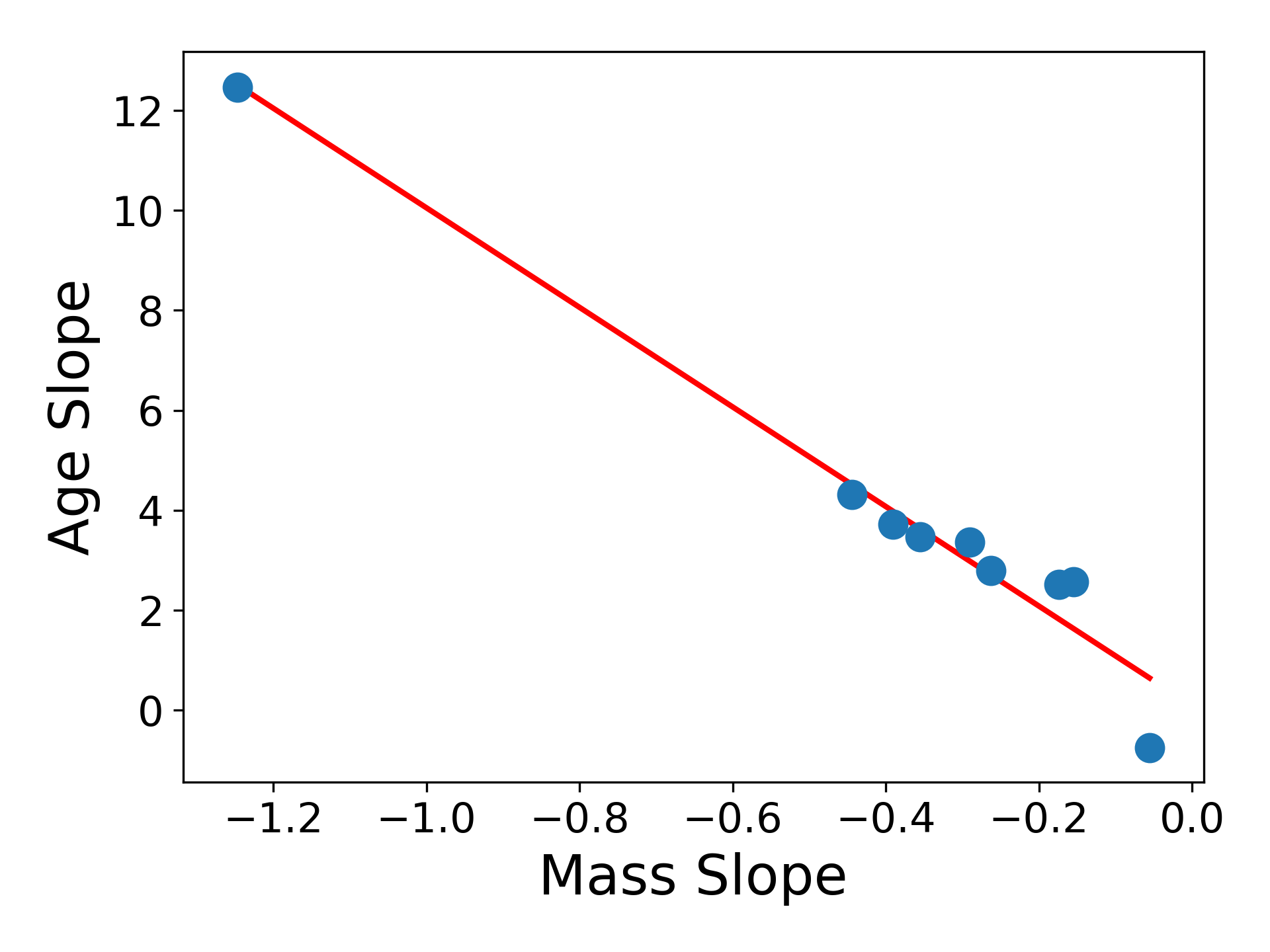}
\caption{The slope in age as a function of the slope in mass for the data presented in Fig.~\ref{fig:individual_grids}.}
\label{fig:individual_mass_slope}
\end{figure}

From Fig.~\ref{fig:individual_grids} we also see an interesting trend where as the overshoot value of the grid increases, the inferred value of the mass and radius decrease. This of course means that the inferred age increases. We can compare the slope of the change between the overshoot subgrids for mass and age. In other words, does a star whose mass is very sensitive to the overshoot value of the subgrid also have an age that is more dependent on the overshoot as well. From the red line of best fit in Fig.~\ref{fig:individual_grids} we can plot the slope of the age change as a function of the slope of the mass change. Figure~\ref{fig:individual_mass_slope} shows this relationship. As expected, there is a trend between the age slope and the mass slope. A star whose mass is affected more by the subgrid's overshoot value also has an age that is more impacted by the overshoot value. 

\subsection{Adiabatic Overshoot vs. Overmixing}
\label{sec:model_physics}
\begin{figure}
\epsscale{1.0}
\plotone{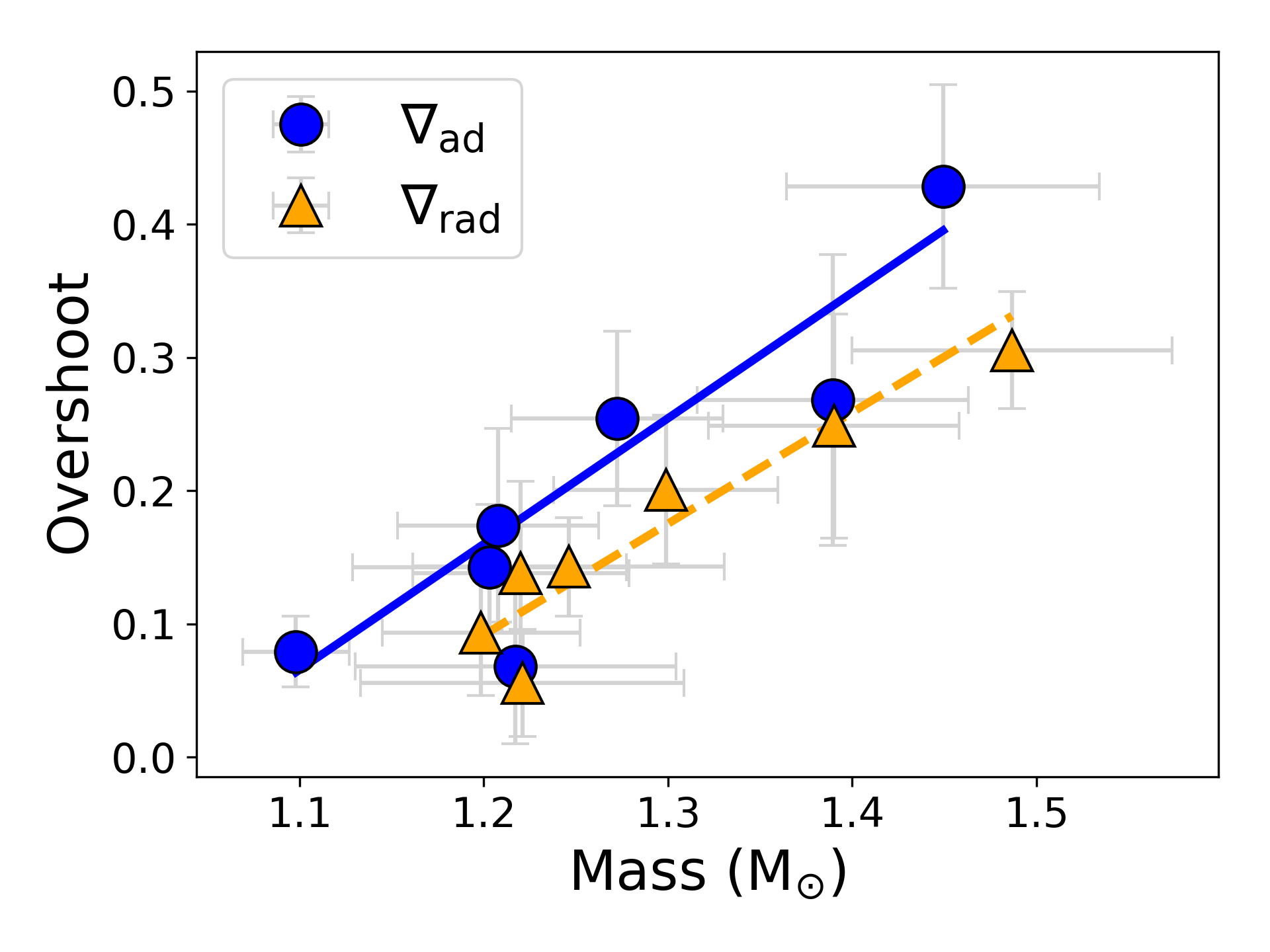}
\caption{Effective overshoot as a function of mass for the stars in the sample with an adiabatic overshooting region (blue points) and with a radiative temperature gradient in the overshoot region (orange points). Overshoot is in units of $H_P$.}
\label{fig:grad_no_grad}
\end{figure}

When constructing the stellar models with overshoot, a choice must be made regarding the temperature gradient in the overshooting region. In the models thus far, the overshooting region has an adiabatic temperature gradient. For a subset of the stars we also create grids where the overshoot region has a radiative gradient, but the composition is uniform and the same as the core, also known as overmixing. \cite{Zahn1991} refers to the use of the radiative gradient as ``overshooting'' and calls the use of the adiabatic gradient ``penetrative convection''. The different overshoot implementations will change the model frequencies, and thus impact the likelihood weighted averages of the inferred stellar properties. Figure~\ref{fig:grad_no_grad} compares the relationship between overshoot and mass for these two overshoot implementations. As can be seen in Fig.~\ref{fig:grad_no_grad}, there is a positive trend between overshoot and mass regardless of the temperature gradient used. The strength of the trend is fairly similar, with a weighted line of best fit slope of 0.94 and 0.84 for the adiabatic and radiative gradients respectively.

Since the overshoot implementation is different, we also expect to see some level of difference in the inferred masses of the stars. The masses derived from the two grids are overall in good agreement, with a mean absolute difference of 2.7\%. With the exception of one star, the differences in masses between the two grids are all within the $1 \sigma$ uncertainty. Even for the star with the largest mass difference, the change is less than 10\%.   

$ $

\subsection{Overshoot as a Function of Core Size}
\begin{figure*}
\epsscale{1.0}
\plotone{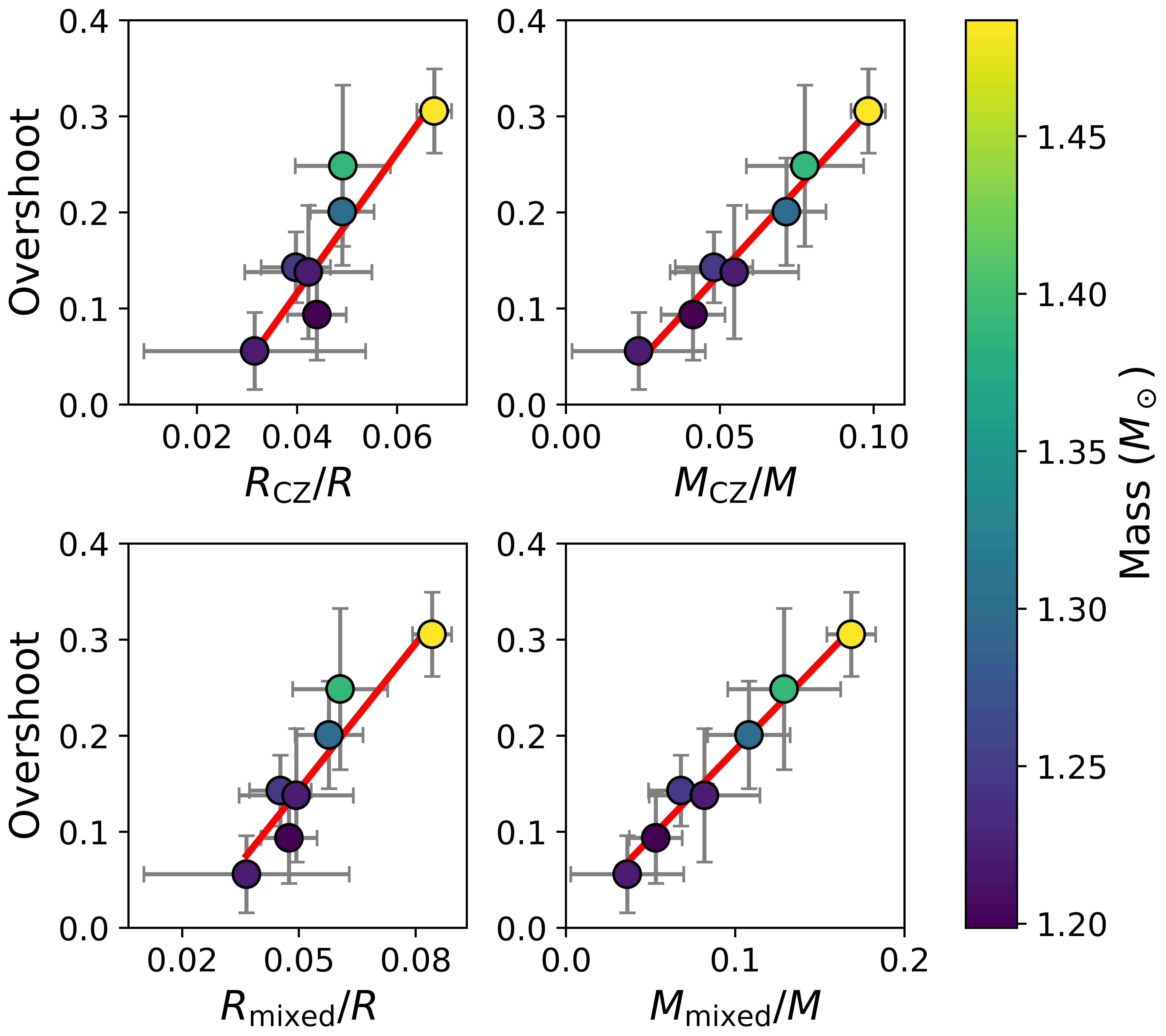}
\caption{The effective overshoot value (in units of $H_P$) plotted as a function of fractional core radius and mass. From the top left, the panels are: convective core radius fraction ($R_\mathrm{CZ} / R$), convective core mass fraction ($M_\mathrm{CZ} / M$), mixed core radius fraction ($R_\mathrm{mixed} / R$), and mixed core mass fraction ($M_\mathrm{mixed} / M$). The color of the points indicate the star's total mass. The red line is a weighted linear best fit. The inferred values for this plot come from the grid of models that implement a radiative gradient in the convective region (overmixing).}
\label{fig:zone_sizes}
\end{figure*}

We also investigate the relationship between overshoot and the mass of the convective core. Using the grid of models with the overmixing convection implementation (see Sec.~\ref{sec:model_physics}), we determine the inferred core mass and core size for the stars. Figure~\ref{fig:zone_sizes} plots overshoot as a function of the convective core mass fraction ($M_\mathrm{CZ} / M$), convective core radius fraction ($R_\mathrm{CZ} / R$), mixed core mass fraction ($M_\mathrm{mixed} / M$), and mixed core radius fraction ($R_\mathrm{mixed} / R$). The convective core fraction refers to just the core, while the mixed core fraction refers to the convective core and the overshoot region together. Clearly, there is a positive trend in all four panels, meaning that stars with larger convective cores have more overshoot. The trend between overshoot and core mass, for both the core only and the full mixed region, has very little scatter and the points nearly all fall directly on the weighted line of best fit. Therefore, the core mass fraction appears to be an excellent indicator of overshoot, at least for this sample of stars.

\subsection{Using the $r_{010}$ Ratios as a Predictor of Overshoot}
A previous study by \cite{Deheuvels2016} investigated the possibility of constraining overshoot by fitting a polynomial to the $r_{010}$ ratios. Since the mean value and the slope of the $r_{010}$ ratios are related to the size of the core and the mixing beyond the core \citep[see, e.g.,][]{Popielski2005,Deheuvels2010,Aguirre2011} then the form of the dependence of $r_{010}$ against frequency on overshoot could be used to estimate the amount of overshoot. \cite{Deheuvels2016} used a 2nd order polynomial to express the dependence with a polynomial of the form:
\begin{equation}
P(\nu) = a_0 + a_1(\nu-\beta) + a_2(\nu-\gamma_1)(\nu-\gamma_2),
\label{eq:polynomial}
\end{equation} 
where $\beta$, $\gamma_1$, and $\gamma_2$ ensure that the polynomial components are orthogonal. When plotting models of different overshoot amounts in the $(a_1,a_0)$ plane \cite{Deheuvels2016} found that the stars tend to separate based on the overshoot value.

We perform a similar study with our model stars. For each model we first calculate the $r_{010}$ ratios, where we use the simplest form of the ratios \cite[see, e.g.,][]{BasuBook2017} defining $r_{01}$ and $r_{10}$ as,
\begin{equation}
r_{01}(n)= \frac{\nu_{n,0} - \frac{1}{2}(\nu_{n-1,1} + \nu_{n,1})}{\nu_{n,1} - \nu_{n-1,1}}
\end{equation} 
and
\begin{equation}
\label{eq:r10}
r_{10}(n)= \frac{\frac{1}{2}(\nu_{n,0}+\nu_{n+1,0})-\nu_{n,1}}{\nu_{n+1,0} - \nu_{n,0}}.
\end{equation} 

\begin{figure*}
\epsscale{1.0}
\plotone{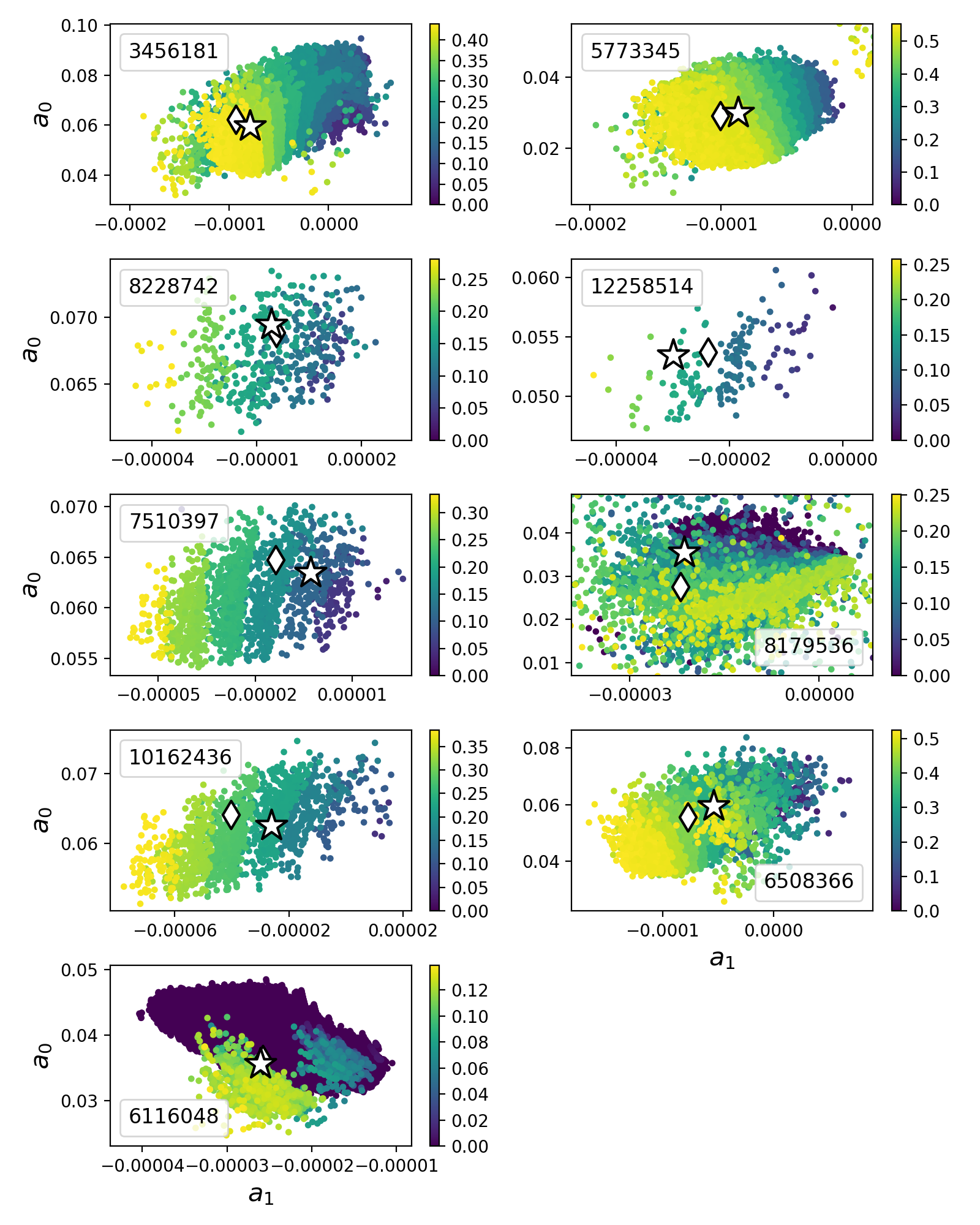}
\caption{The stellar models plotted in the ($a_1$,$a_0$) plane. Each panel represents one of the stars in the sample, labeled by KIC number. The colors of the points indicate the effective overshoot value, in units of $H_P$. The white star symbol shows the observed data and the white diamond point is the best fit model.}
\label{fig:a0a1_all_stars}
\end{figure*}

The polynomial from Eq.~\ref{eq:polynomial} is then fit to the $r_{010}$ ratios of the stellar models as well as to the observed data. For the polynomial fit to the model stars, we only include the model frequencies that are within the observed frequency range for that star. We then examine the models plotted in the $(a_1,a_0)$ plane, along with the location in the plane for the observed data. Figure~\ref{fig:a0a1_all_stars} shows the $(a_1,a_0)$ plane for each star in the sample. The points are colored according to their effective overshoot values. The white points indicate the location of the best-fit model and the observed data. Note that only models with likelihoods of more than 1\% of the best-fit model's likelihood are included in the figure. This is done to exclude the worst fitting models. It can be seen that models with different overshoot values tend to spread themselves out across the $(a_1,a_0)$ plane. While there is overlap between the overshoot values, a model's location in the $(a_1,a_0)$ plane does seem to have some dependence on overshoot. Note that the panel for KIC 6116048 has an abundance of stars with an overshoot of 0 (purple points) and the trend with overshoot and location in the $(a_1,a_0)$ plane is less clear. This is the lowest mass star in the sample, and as a result many of the models did not have a convective core. In fact, 95\% of the models for this star had no convective core. Since there is no convective core in these models, there cannot be any overshoot and we have defined this to be an effective overshoot of 0.

There are two main differences in our approach to determining the $a_0$ and $a_1$ coefficients compared to \cite{Deheuvels2016}. First of all, we make use of the simpler form of the $r_{01}$ and $r_{10}$ ratios. Additionally, when fitting the polynomial, \cite{Deheuvels2016} first fit the observed frequencies and used those results to determine $\beta$, $\gamma_1$, and $\gamma_2$. The value of $\beta$, $\gamma_1$, and $\gamma_2$ is then held fixed for the model stars. We, however, allow $\beta$, $\gamma_1$, and $\gamma_2$ to be free parameters for the model stars. Thus, for each model we are allowing all six coefficients to be free. The reason that \cite{Deheuvels2016} fixed the orthogonal coefficients was so that when comparing between the models and the observed star, the $a_n$ coefficients could be compared more directly since the rest of the polynomial was identical. We tested holding the orthogonal coefficients fixed as opposed to allowing them to be free parameters for several of our stars and saw no major difference in the results of the $a_0$ and $a_1$ values. 

\begin{figure*}
\epsscale{1.0}
\plotone{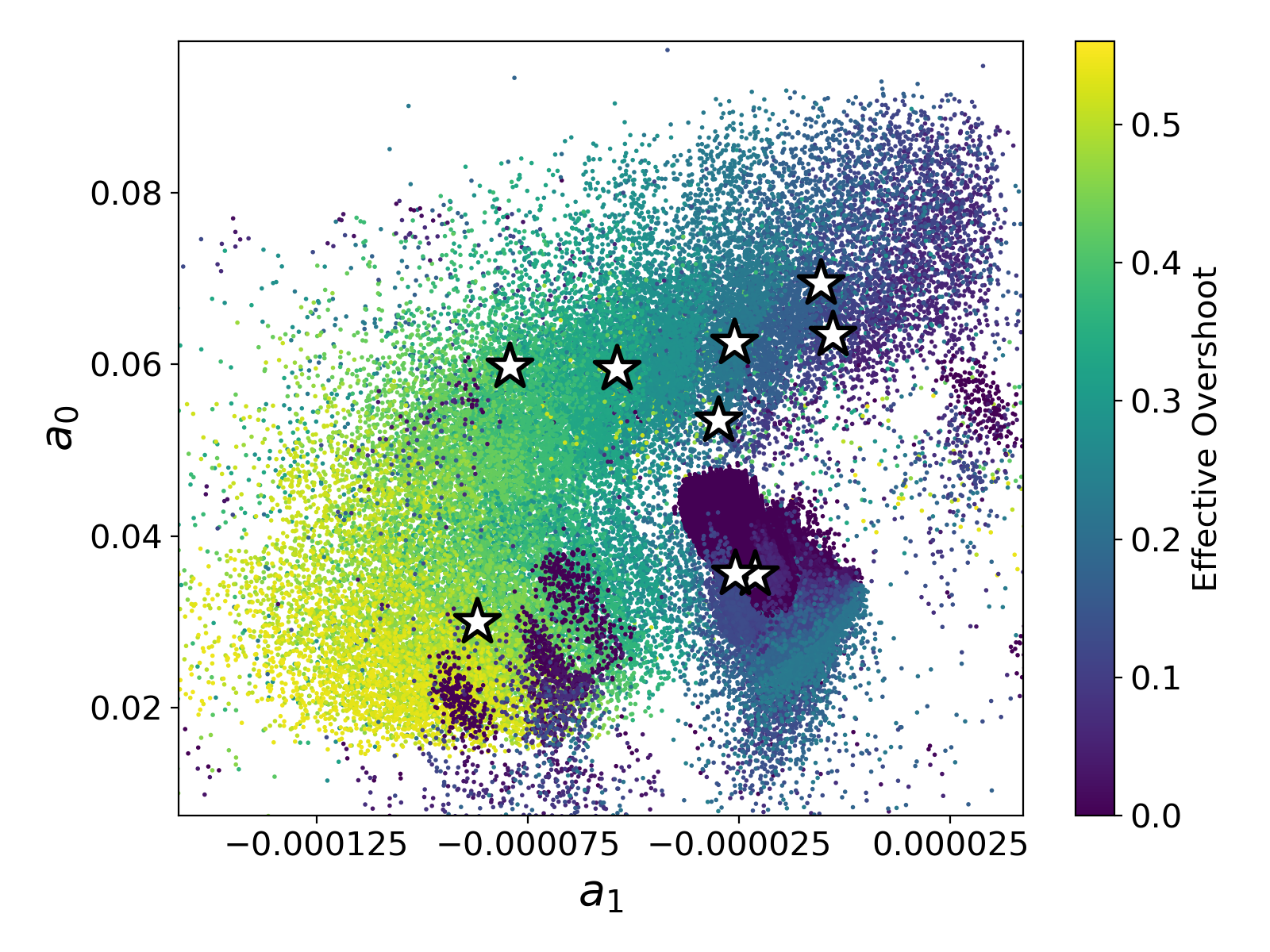}
\caption{The stellar models in the ($a_1$,$a_0$) plane for all 9 stars. Colors indicate the effective overshoot value of the model, in units of $H_P$. The white symbol is the observed value.}
\label{fig:a1a0_one_panel}
\end{figure*}

The advantage of allowing all the coefficients to be free, as opposed to based on the observed star, is that we can then compare all our models for all the stars in the same $(a_1,a_0)$ plane as an ensemble. This can be seen in Figure~\ref{fig:a1a0_one_panel}. Here, we see that the models spread themselves out in the $(a_1,a_0)$ plane roughly according to their overshoot, even when using the ensemble of stars, and not just one set of grids at a time. There is an island of points (in the lower right) of very low overshoot that do not quite follow the trend as clearly. These points are from the two stars with the lowest inferred overshoot and, as mentioned previously, a large portion of the models for these stars did not have a convective core. For the remained of the stars however, we see a fairly continuous and smooth trend between overshoot and location in the ($a_1$,$a_0$) plane. 

It is of interest that in both Fig.~\ref{fig:a0a1_all_stars} and~\ref{fig:a1a0_one_panel} we see most of the overshoot separation occurs in the $a_1$ coefficient. In \cite{Deheuvels2016} the separation was occurring sometimes in the $a_1$ direction, sometimes in $a_0$, and in some cases a combination of both. These results indicate that for this set of stars, the $a_1$ coefficient may perhaps be used as an indicator of overshoot amount. 

\begin{figure}
\epsscale{1.0}
\plotone{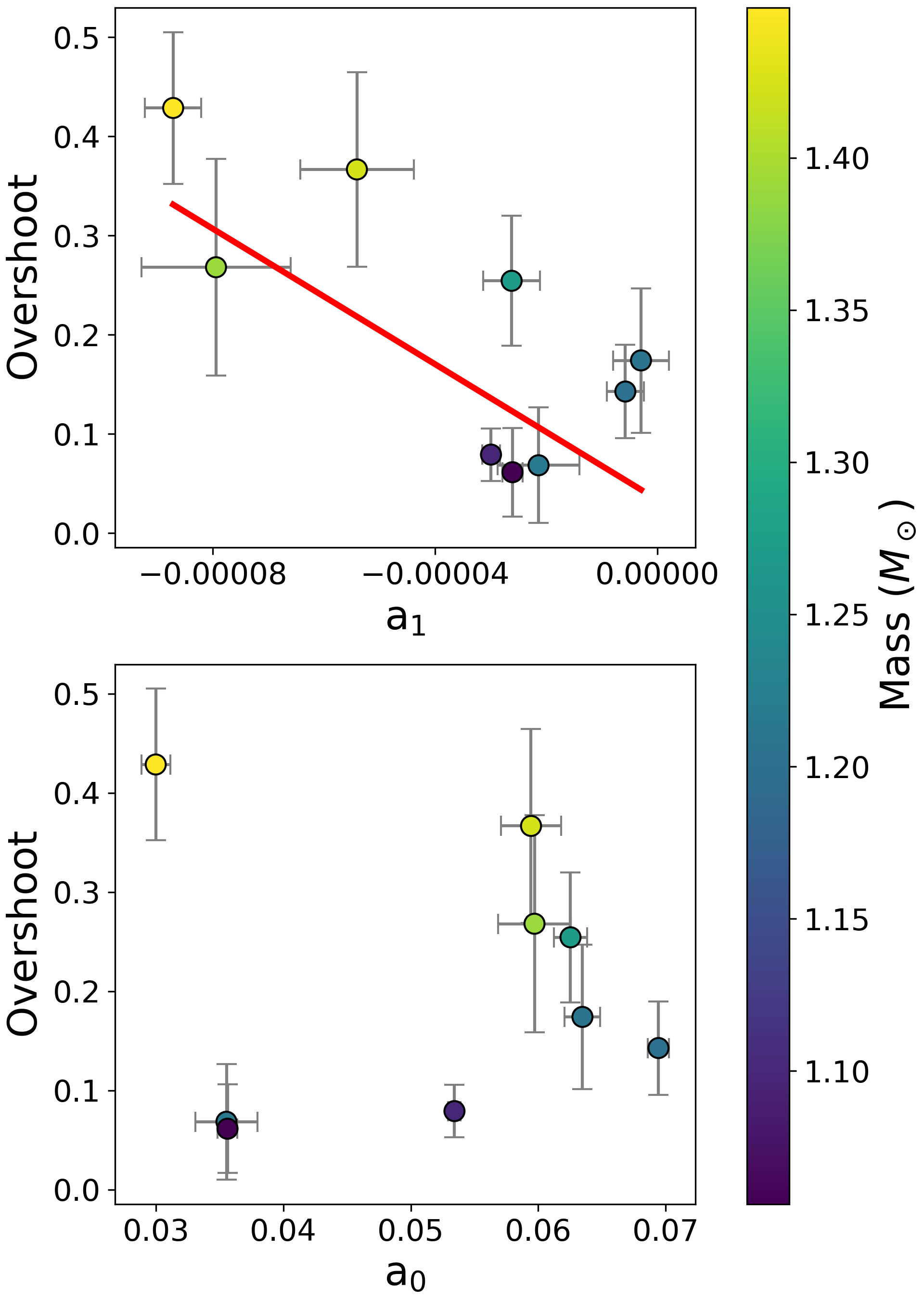}
\caption{\textbf{Top:} Effective overshoot as a function of the $a_1$ coefficient for the sample of stars. Here the $a_1$ value is determined from the observed frequencies of the stars. The red line is a weighted line of best fit. \textbf{Bottom:} Effective overshoot as a function of the $a_0$ coefficient for the sample of stars. In both panels overshoot is in units of $H_P$. See Fig.~\ref{fig:a1a0_vs_input_overshoot} in the Appendix for a version of this figure using the input overshoot.}
\label{fig:a1a0_vs_overshoot}
\end{figure}

In Figure~\ref{fig:a1a0_vs_overshoot} we plot overshoot as a function of $a_1$ and $a_0$ for the observed frequencies for all the stars in the sample. Here we see a negative trend between $a_1$ and overshoot, albeit with some scatter. A possible trend between $a_1$ and overshoot would be extremely useful in regards to constraining the overshoot value. If a star's observed frequencies are known, then $a_1$ can be quickly obtained and a rough estimate of overshoot could be made. While we see a more clear and tighter trend between mass and overshoot, determining the mass of a star either requires generating a lot of models, or relying on the asteroseismic scaling relations. Hence, using the $a_1$ coefficient may possibly be a quicker method, provided that the observed frequencies are known. However, more investigation into the $a_1$--overshoot relation is needed. The trend in Fig~\ref{fig:a1a0_vs_overshoot} does contain a lot of scatter. We do not see a trend between $a_0$ and overshoot. For a version of Fig.~\ref{fig:a1a0_vs_overshoot} created using the input overshoot instead of the effective overshoot, see Fig.~\ref{fig:a1a0_vs_input_overshoot} in the Appendix.    

\begin{figure}
\epsscale{1.0}
\plotone{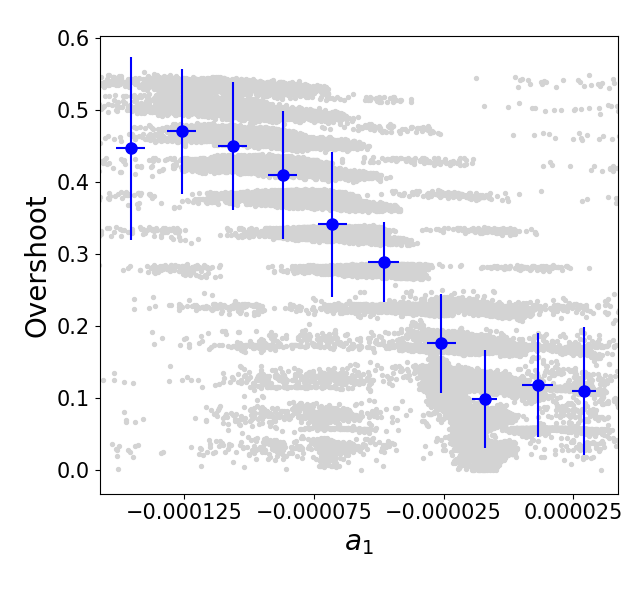}
\caption{Effective overshoot (in units of $H_P$) as a function of the $a_1$ coefficient for stellar models. The models are plotted in gray and the blue points are the mean of the binned data.}
\label{fig:a1a0_vs_overshoot_models}
\end{figure}

We also investigate the trend between $a_1$ and overshoot in the stellar models themselves. Figure~\ref{fig:a1a0_vs_overshoot_models} examines overshoot as a function of $a_1$ using the models from Fig.~\ref{fig:a1a0_one_panel}. The gray points, which are the individual models, show that a given $a_1$ value can correspond to a range of overshoot amounts. However, when we bin the data and plot the mean of each bin (blue points) a potential trend between overshoot and $a_1$ is seen. The models are grouped by $a_1$ into 10 bins. For each bin the blue point represents the mean overshoot and $a_1$ value for that bin, with the errorbars being the standard deviation within the bin. From the binned data there is an apparent negative trend between overshoot and $a_1$ in the models. So, not only do we see a trend between overshoot and $a_1$ for the inferred properties of the observed star, we also see this trend for the models themselves.

\section{Summary and Conclusions}
\label{sec:conclusion}
In this work we determined the amount of overshoot needed to properly model a set of stars from the \textit{Kepler} Asteroseismic LEGACY Sample. Searching for trends between overshoot and stellar properties, we find a strong positive relationship between overshoot and stellar mass for the mass range of our sample (1.1-1.5$M_\odot$). This trend with mass is present regardless of whether the temperature gradient in the overshooting region is adiabatic or radiative. We additionally highlight the important difference between a stellar model's input overshoot and the resulting effective overshoot value. 

The trend between mass and overshoot holds promising potential to help reduce the ambiguity and uncertainty in the overshoot parameter. If a rough estimate of the overshoot amount can be determined using the star's mass, then the parameter space of overshoot values needed to be included in the models can be made smaller. While there is still a need to include a range of overshoot values in the models, this mass--overshoot relationship can help narrow down the range of the overshoot parameter.

Additionally, we examined the impact that the overshoot value can have on inferred stellar properties. By using only one overshoot subgrid at a time, we determine how the derived stellar parameters change based on the overshoot amount used. We see changes in the inferred stellar mass, radius, and age can be as large as 14\%, 6\%, and 50\% respectively. These differences highlight the sensitivity of inferred stellar properties to overshoot, and show the importance of selecting the correct overshoot value (or range of values) for models. Since the inferred properties are so sensitive to the overshoot amount, these results strongly suggest that including a range of overshoot amounts is preferable to assuming a singe fixed value. Many current model libraries that implement overshoot use a single fixed value of overshoot above some certain critical mass, with the overshoot amount decreasing for lower masses \citep{Demarque2004, Pietrinferni2004, Bressan2012, Hidalgo2018}.

Finally, we investigated using the $r_{010}$ ratios as an indicator of overshoot extent. By fitting a polynomial to the $r_{010}$ ratios we see that by plotting the models in the ($a_1,a_0$) plane the models generally spread themselves out based on the overshoot amount. We see most of this separation in the direction of the $a_1$ coefficient. For this set of stars we see a negative trend between overshoot and $a_1$. While this relationship between $a_1$ and overshoot has a decent amount of scatter, this preliminary work shows potential that the $a_1$ value may also be used to constrain the overshoot amount. Certainly the relationship between $a_1$ and overshoot merits further investigation.

In summary, based on these results, we see a promising potential towards using mass and/or the $a_1$ coefficient as a predictor of overshoot. While the relationships between overshoot and mass and overshoot and $a_1$ have scatter, they still will be very useful in helping to reduce the ambiguity and uncertainty in the overshoot value.  While there may always be a need to include a range in overshoot values, based on the demonstrated sensitivity of the inferred stellar parameters to overshoot, obtaining an estimate of the overshoot value using mass or $a_1$ will help constrain the possible values of overshoot. Future work towards a better understanding of the relationship between mass, $a_1$, and overshoot will include increasing the number of stars in the sample and increasing the mass range as well.

\acknowledgments This work was partially supported by NSF grant AST-1514676 and NASA grant NNX16AI09G to SB.

\software{YREC \citep{Demarque2008}}

\begin{figure*}
\epsscale{1.0}
\plotone{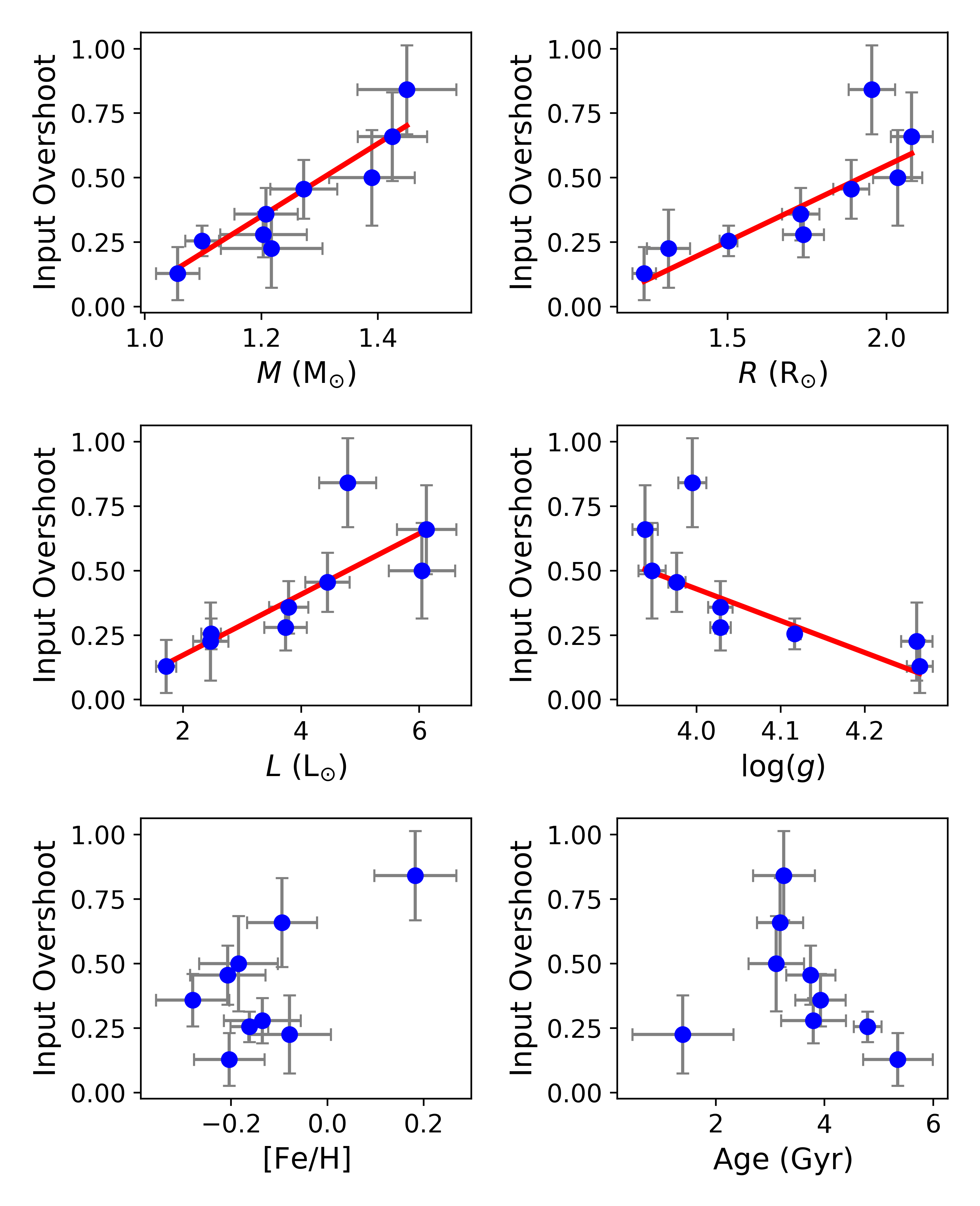}
\caption{The likelihood weighted average of the input overshoot value (in units of $H_P$) plotted as a function of a variety of stellar properties for the stars in the study. The red line is a weighted linear best fit.}
\label{fig:input_overshoot_multiplot}
\end{figure*}

\begin{figure}
\epsscale{0.6}
\plotone{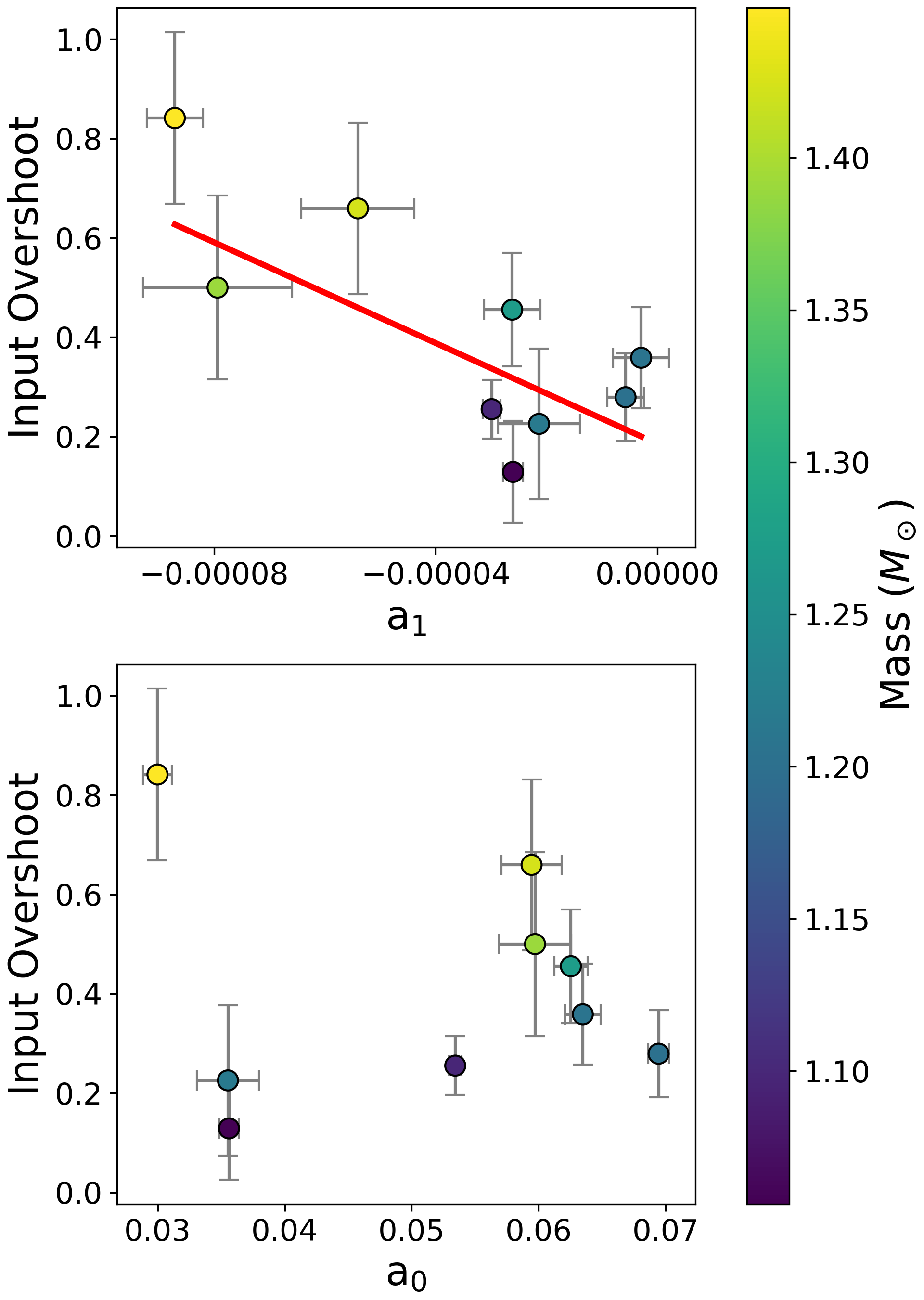}
\caption{\textbf{Top:} Input overshoot as a function of the $a_1$ coefficient for the sample of stars. Here the $a_1$ value is determined from the observed frequencies of the stars. The red line is a weighted line of best fit. \textbf{Bottom:} Input overshoot as a function of the $a_0$ coefficient for the sample of stars. In both panels overshoot is in units of $H_P$.}
\label{fig:a1a0_vs_input_overshoot}
\end{figure}

\appendix
\label{sec:Appendix}
Since the value of the input overshoot is what is supplied to the code when constructing models, it is also of interest to reexamine the main results (Figures~\ref{fig:overshoot_multiplot} and \ref{fig:a1a0_vs_overshoot}) using the input overshoot instead of the effective overshoot. As can be seen in Figures \ref{fig:input_overshoot_multiplot} and \ref{fig:a1a0_vs_input_overshoot}, the trends are still present.

\bibliography{overshoot}

\begin{thebibliography}{}
\expandafter\ifx\csname natexlab\endcsname\relax\def\natexlab#1{#1}\fi

\bibitem[{{Adelberger} {et~al.}(1998){Adelberger}, {Austin}, {Bahcall},
  {Balantekin}, {Bogaert}, {Brown}, {Buchmann}, {Cecil}, {Champagne}, {de
  Braeckeleer}, {Duba}, {Elliott}, {Freedman}, {Gai}, {Goldring}, {Gould},
  {Gruzinov}, {Haxton}, {Heeger}, {Henley}, {Johnson}, {Kamionkowski},
  {Kavanagh}, {Koonin}, {Kubodera}, {Langanke}, {Motobayashi}, {Pandharipande},
  {Parker}, {Robertson}, {Rolfs}, {Sawyer}, {Shaviv}, {Shoppa}, {Snover},
  {Swanson}, {Tribble}, {Turck-Chi{\`e}ze}, \& {Wilkerson}}]{Adelberger1998}
{Adelberger}, E.~G., {Austin}, S.~M., {Bahcall}, J.~N., {et~al.} 1998, Reviews
  of Modern Physics, 70, 1265

\bibitem[{{Aerts}(2015)}]{Aerts2015}
{Aerts}, C. 2015, in IAU Symposium, Vol. 307, New Windows on Massive Stars, ed.
  G.~{Meynet}, C.~{Georgy}, J.~{Groh}, \& P.~{Stee}, 154--164

\bibitem[{{Angelou} {et~al.}(2020){Angelou}, {Bellinger}, {Hekker}, {Mints},
  {Elsworth}, {Basu}, \& {Weiss}}]{Angelou2020}
{Angelou}, G.~C., {Bellinger}, E.~P., {Hekker}, S., {et~al.} 2020, \mnras, 493,
  4987

\bibitem[{{Barmina} {et~al.}(2002){Barmina}, {Girardi}, \&
  {Chiosi}}]{Barmina2002}
{Barmina}, R., {Girardi}, L., \& {Chiosi}, C. 2002, \aap, 385, 847

\bibitem[{{Basu} \& {Chaplin}(2017)}]{BasuBook2017}
{Basu}, S., \& {Chaplin}, W.~J. 2017, {Asteroseismic Data Analysis: Foundations
  and Techniques}

\bibitem[{{B{\"o}hm-Vitense}(1958)}]{BohmVitense1958}
{B{\"o}hm-Vitense}, E. 1958, \zap, 46, 108

\bibitem[{{Bressan} {et~al.}(2012){Bressan}, {Marigo}, {Girardi}, {Salasnich},
  {Dal Cero}, {Rubele}, \& {Nanni}}]{Bressan2012}
{Bressan}, A., {Marigo}, P., {Girardi}, L., {et~al.} 2012, \mnras, 427, 127

\bibitem[{{Carraro} {et~al.}(1993){Carraro}, {Bertelli}, {Bressan}, \&
  {Chiosi}}]{Carraro1993}
{Carraro}, G., {Bertelli}, G., {Bressan}, A., \& {Chiosi}, C. 1993, \aaps, 101,
  381

\bibitem[{{Chin} \& {Stothers}(1991)}]{Chin1991}
{Chin}, C.-W., \& {Stothers}, R.~B. 1991, \apjs, 77, 299

\bibitem[{{Claret} \& {Torres}(2016)}]{Claret2016}
{Claret}, A., \& {Torres}, G. 2016, \aap, 592, A15

\bibitem[{{Claret} \& {Torres}(2017)}]{Claret2017}
---. 2017, \apj, 849, 18

\bibitem[{{Claret} \& {Torres}(2018)}]{Claret2018}
---. 2018, \apj, 859, 100

\bibitem[{{Claret} \& {Torres}(2019)}]{Claret2019}
---. 2019, \apj, 876, 134

\bibitem[{{Constantino} \& {Baraffe}(2018)}]{Constantino2018}
{Constantino}, T., \& {Baraffe}, I. 2018, \aap, 618, A177

\bibitem[{{Daniel} {et~al.}(1994){Daniel}, {Latham}, {Mathieu}, \&
  {Twarog}}]{Daniel1994}
{Daniel}, S.~A., {Latham}, D.~W., {Mathieu}, R.~D., \& {Twarog}, B.~A. 1994,
  \pasp, 106, 281

\bibitem[{{Degroote} {et~al.}(2010){Degroote}, {Aerts}, {Baglin}, {Miglio},
  {Briquet}, {Noels}, {Niemczura}, {Montalban}, {Bloemen}, {Oreiro},
  {Vu{\v{c}}kovi{\'c}}, {Smolders}, {Auvergne}, {Baudin}, {Catala}, \&
  {Michel}}]{Degroote2010}
{Degroote}, P., {Aerts}, C., {Baglin}, A., {et~al.} 2010, \nat, 464, 259

\bibitem[{{Deheuvels} {et~al.}(2016){Deheuvels}, {Brand{\~a}o}, {Silva
  Aguirre}, {Ballot}, {Michel}, {Cunha}, {Lebreton}, \&
  {Appourchaux}}]{Deheuvels2016}
{Deheuvels}, S., {Brand{\~a}o}, I., {Silva Aguirre}, V., {et~al.} 2016, \aap,
  589, A93

\bibitem[{{Deheuvels} \& {Michel}(2011)}]{Deheuvels2011}
{Deheuvels}, S., \& {Michel}, E. 2011, \aap, 535, A91

\bibitem[{{Deheuvels} {et~al.}(2010){Deheuvels}, {Michel}, {Goupil}, {Marques},
  {Mosser}, {Dupret}, {Lebreton}, {Pichon}, \& {Morel}}]{Deheuvels2010}
{Deheuvels}, S., {Michel}, E., {Goupil}, M.~J., {et~al.} 2010, \aap, 514, A31

\bibitem[{{Demarque} {et~al.}(2008){Demarque}, {Guenther}, {Li}, {Mazumdar}, \&
  {Straka}}]{Demarque2008}
{Demarque}, P., {Guenther}, D.~B., {Li}, L.~H., {Mazumdar}, A., \& {Straka},
  C.~W. 2008, \apss, 316, 31

\bibitem[{{Demarque} {et~al.}(1994){Demarque}, {Sarajedini}, \&
  {Guo}}]{Demarque1994}
{Demarque}, P., {Sarajedini}, A., \& {Guo}, X.~J. 1994, \apj, 426, 165

\bibitem[{{Demarque} {et~al.}(2004){Demarque}, {Woo}, {Kim}, \&
  {Yi}}]{Demarque2004}
{Demarque}, P., {Woo}, J.-H., {Kim}, Y.-C., \& {Yi}, S.~K. 2004, \apjs, 155,
  667

\bibitem[{{Ferguson} {et~al.}(2005){Ferguson}, {Alexander}, {Allard}, {Barman},
  {Bodnarik}, {Hauschildt}, {Heffner-Wong}, \& {Tamanai}}]{Ferguson2005}
{Ferguson}, J.~W., {Alexander}, D.~R., {Allard}, F., {et~al.} 2005, \apj, 623,
  585

\bibitem[{{Formicola} {et~al.}(2004){Formicola}, {Imbriani}, {Costantini},
  {Angulo}, {Bemmerer}, {Bonetti}, {Broggini}, {Corvisiero}, {Cruz},
  {Descouvemont}, {F{\"u}l{\"o}p}, {Gervino}, {Guglielmetti}, {Gustavino},
  {Gy{\"u}rky}, {Jesus}, {Junker}, {Lemut}, {Menegazzo}, {Prati}, {Roca},
  {Rolfs}, {Romano}, {Rossi Alvarez}, {Sch{\"u}mann}, {Somorjai}, {Straniero},
  {Strieder}, {Terrasi}, {Trautvetter}, {Vomiero}, \&
  {Zavatarelli}}]{Formicola2004}
{Formicola}, A., {Imbriani}, G., {Costantini}, H., {et~al.} 2004, Physics
  Letters B, 591, 61

\bibitem[{{Grevesse} \& {Sauval}(1998)}]{GS1998}
{Grevesse}, N., \& {Sauval}, A.~J. 1998, \ssr, 85, 161

\bibitem[{{Guenther} {et~al.}(2014){Guenther}, {Demarque}, \&
  {Gruberbauer}}]{Guenther2014}
{Guenther}, D.~B., {Demarque}, P., \& {Gruberbauer}, M. 2014, \apj, 787, 164

\bibitem[{{Hidalgo} {et~al.}(2018){Hidalgo}, {Pietrinferni}, {Cassisi},
  {Salaris}, {Mucciarelli}, {Savino}, {Aparicio}, {Silva Aguirre}, \&
  {Verma}}]{Hidalgo2018}
{Hidalgo}, S.~L., {Pietrinferni}, A., {Cassisi}, S., {et~al.} 2018, \apj, 856,
  125

\bibitem[{{Iglesias} \& {Rogers}(1996)}]{Iglesias1996}
{Iglesias}, C.~A., \& {Rogers}, F.~J. 1996, \apj, 464, 943

\bibitem[{{Kippenhahn} {et~al.}(2012){Kippenhahn}, {Weigert}, \&
  {Weiss}}]{Kippenhahn2012}
{Kippenhahn}, R., {Weigert}, A., \& {Weiss}, A. 2012, {Stellar Structure and
  Evolution}, doi:10.1007/978-3-642-30304-3

\bibitem[{{Kozhurina-Platais} {et~al.}(1997){Kozhurina-Platais}, {Demarque},
  {Platais}, {Orosz}, \& {Barnes}}]{Kozhurina1997}
{Kozhurina-Platais}, V., {Demarque}, P., {Platais}, I., {Orosz}, J.~A., \&
  {Barnes}, S. 1997, \aj, 113, 1045

\bibitem[{{Lebreton} \& {Goupil}(2014)}]{Lebreton2014}
{Lebreton}, Y., \& {Goupil}, M.~J. 2014, \aap, 569, A21

\bibitem[{{Lund} {et~al.}(2017){Lund}, {Silva Aguirre}, {Davies}, {Chaplin},
  {Christensen-Dalsgaard}, {Houdek}, {White}, {Bedding}, {Ball}, {Huber},
  {Antia}, {Lebreton}, {Latham}, {Handberg}, {Verma}, {Basu}, {Casagrande},
  {Justesen}, {Kjeldsen}, \& {Mosumgaard}}]{Lund2017}
{Lund}, M.~N., {Silva Aguirre}, V., {Davies}, G.~R., {et~al.} 2017, \apj, 835,
  172

\bibitem[{{Maeder}(1976)}]{Maeder1976}
{Maeder}, A. 1976, \aap, 47, 389

\bibitem[{{Maeder} \& {Mermilliod}(1981)}]{Maeder1981}
{Maeder}, A., \& {Mermilliod}, J.~C. 1981, \aap, 93, 136

\bibitem[{{Maeder} \& {Meynet}(1989)}]{Maeder1989}
{Maeder}, A., \& {Meynet}, G. 1989, \aap, 210, 155

\bibitem[{{Montalb{\'a}n} {et~al.}(2013){Montalb{\'a}n}, {Miglio}, {Noels},
  {Dupret}, {Scuflaire}, \& {Ventura}}]{Montalban2013}
{Montalb{\'a}n}, J., {Miglio}, A., {Noels}, A., {et~al.} 2013, \apj, 766, 118

\bibitem[{{Moravveji} {et~al.}(2015){Moravveji}, {Aerts}, {P{\'a}pics},
  {Triana}, \& {Vandoren}}]{Moravveji2015}
{Moravveji}, E., {Aerts}, C., {P{\'a}pics}, P.~I., {Triana}, S.~A., \&
  {Vandoren}, B. 2015, \aap, 580, A27

\bibitem[{{Neiner} {et~al.}(2012){Neiner}, {Mathis}, {Saio}, {Lovekin},
  {Eggenberger}, \& {Lee}}]{Neiner2012}
{Neiner}, C., {Mathis}, S., {Saio}, H., {et~al.} 2012, \aap, 539, A90

\bibitem[{{Nordstroem} {et~al.}(1997){Nordstroem}, {Andersen}, \&
  {Andersen}}]{Nordstroem1997}
{Nordstroem}, B., {Andersen}, J., \& {Andersen}, M.~I. 1997, \aap, 322, 460

\bibitem[{{Paxton} {et~al.}(2018){Paxton}, {Schwab}, {Bauer}, {Bildsten},
  {Blinnikov}, {Duffell}, {Farmer}, {Goldberg}, {Marchant}, {Sorokina},
  {Thoul}, {Townsend}, \& {Timmes}}]{Paxton2018}
{Paxton}, B., {Schwab}, J., {Bauer}, E.~B., {et~al.} 2018, \apjs, 234, 34

\bibitem[{{Pietrinferni} {et~al.}(2004){Pietrinferni}, {Cassisi}, {Salaris}, \&
  {Castelli}}]{Pietrinferni2004}
{Pietrinferni}, A., {Cassisi}, S., {Salaris}, M., \& {Castelli}, F. 2004, \apj,
  612, 168

\bibitem[{{Popielski} \& {Dziembowski}(2005)}]{Popielski2005}
{Popielski}, B.~L., \& {Dziembowski}, W.~A. 2005, \actaa, 55, 177

\bibitem[{{Prather} \& {Demarque}(1974)}]{Prather1974}
{Prather}, M.~J., \& {Demarque}, P. 1974, \apj, 193, 109

\bibitem[{{Reese}(2016)}]{Reese2016}
{Reese}, D.~R. 2016, {AIMS: Asteroseismic Inference on a Massive Scale}, , ,
  ascl:1611.014

\bibitem[{{Rendle} {et~al.}(2019){Rendle}, {Buldgen}, {Miglio}, {Reese},
  {Noels}, {Davies}, {Campante}, {Chaplin}, {Lund}, {Kuszlewicz}, {Scott},
  {Scuflaire}, {Ball}, {Smetana}, \& {Nsamba}}]{Rendle2019}
{Rendle}, B.~M., {Buldgen}, G., {Miglio}, A., {et~al.} 2019, \mnras, 484, 771

\bibitem[{{Rogers} \& {Nayfonov}(2002)}]{Rogers2002}
{Rogers}, F.~J., \& {Nayfonov}, A. 2002, \apj, 576, 1064

\bibitem[{{Roxburgh}(1992)}]{Roxburgh1992}
{Roxburgh}, I.~W. 1992, \aap, 266, 291

\bibitem[{{Roxburgh} \& {Vorontsov}(2003)}]{Roxburgh2003}
{Roxburgh}, I.~W., \& {Vorontsov}, S.~V. 2003, \aap, 411, 215

\bibitem[{{Salaris} \& {Cassisi}(2017)}]{Salaris2017}
{Salaris}, M., \& {Cassisi}, S. 2017, Royal Society Open Science, 4, 170192

\bibitem[{{Silva Aguirre} {et~al.}(2011){Silva Aguirre}, {Ballot}, {Serenelli},
  \& {Weiss}}]{Aguirre2011}
{Silva Aguirre}, V., {Ballot}, J., {Serenelli}, A.~M., \& {Weiss}, A. 2011,
  \aap, 529, A63

\bibitem[{{Silva Aguirre} {et~al.}(2013){Silva Aguirre}, {Basu}, {Brand{\~a}o},
  {Christensen-Dalsgaard}, {Deheuvels}, {Do{\u{g}}an}, {Metcalfe}, {Serenelli},
  {Ballot}, {Chaplin}, {Cunha}, {Weiss}, {Appourchaux}, {Casagrande},
  {Cassisi}, {Creevey}, {Garc{\'\i}a}, {Lebreton}, {Noels}, {Sousa}, {Stello},
  {White}, {Kawaler}, \& {Kjeldsen}}]{Aguirre2013}
{Silva Aguirre}, V., {Basu}, S., {Brand{\~a}o}, I.~M., {et~al.} 2013, \apj,
  769, 141

\bibitem[{{Silva Aguirre} {et~al.}(2015){Silva Aguirre}, {Davies}, {Basu},
  {Christensen-Dalsgaard}, {Creevey}, {Metcalfe}, {Bedding}, {Casagrande},
  {Handberg}, {Lund}, {Nissen}, {Chaplin}, {Huber}, {Serenelli}, {Stello}, {Van
  Eylen}, {Campante}, {Elsworth}, {Gilliland}, {Hekker}, {Karoff}, {Kawaler},
  {Kjeldsen}, \& {Lundkvist}}]{Aguirre2015}
{Silva Aguirre}, V., {Davies}, G.~R., {Basu}, S., {et~al.} 2015, \mnras, 452,
  2127

\bibitem[{{Silva Aguirre} {et~al.}(2017){Silva Aguirre}, {Lund}, {Antia},
  {Ball}, {Basu}, {Christensen-Dalsgaard}, {Lebreton}, {Reese}, {Verma},
  {Casagrande}, {Justesen}, {Mosumgaard}, {Chaplin}, {Bedding}, {Davies},
  {Handberg}, {Houdek}, {Huber}, {Kjeldsen}, {Latham}, {White}, {Coelho},
  {Miglio}, \& {Rendle}}]{Aguirre2017}
{Silva Aguirre}, V., {Lund}, M.~N., {Antia}, H.~M., {et~al.} 2017, \apj, 835,
  173

\bibitem[{{Stancliffe} {et~al.}(2015){Stancliffe}, {Fossati}, {Passy}, \&
  {Schneider}}]{Stancliffe2015}
{Stancliffe}, R.~J., {Fossati}, L., {Passy}, J.~C., \& {Schneider}, F.~R.~N.
  2015, \aap, 575, A117

\bibitem[{{Stothers}(1991)}]{Stothers1991}
{Stothers}, R.~B. 1991, \apj, 383, 820

\bibitem[{{Valle} {et~al.}(2016){Valle}, {Dell'Omodarme}, {Prada Moroni}, \&
  {Degl'Innocenti}}]{Valle2016}
{Valle}, G., {Dell'Omodarme}, M., {Prada Moroni}, P.~G., \& {Degl'Innocenti},
  S. 2016, \aap, 587, A16

\bibitem[{{Valle} {et~al.}(2017){Valle}, {Dell'Omodarme}, {Prada Moroni}, \&
  {Degl'Innocenti}}]{Valle2017}
---. 2017, \aap, 600, A41

\bibitem[{{Valle} {et~al.}(2018){Valle}, {Dell'Omodarme}, {Prada Moroni}, \&
  {Degl'Innocenti}}]{Valle2018}
---. 2018, \aap, 615, A62

\bibitem[{{VandenBerg} {et~al.}(2007){VandenBerg}, {Gustafsson}, {Edvardsson},
  {Eriksson}, \& {Ferguson}}]{VandenBerg2007}
{VandenBerg}, D.~A., {Gustafsson}, B., {Edvardsson}, B., {Eriksson}, K., \&
  {Ferguson}, J. 2007, \apjl, 666, L105

\bibitem[{{Viani} {et~al.}(2018){Viani}, {Basu}, {Joel Ong J.}, {Bonaca}, \&
  {Chaplin}}]{Viani2018}
{Viani}, L.~S., {Basu}, S., {Joel Ong J.}, M., {Bonaca}, A., \& {Chaplin},
  W.~J. 2018, \apj, 858, 28

\bibitem[{{Weiss} \& {Schlattl}(2008)}]{Weiss2008}
{Weiss}, A., \& {Schlattl}, H. 2008, \apss, 316, 99

\bibitem[{{Woo} \& {Demarque}(2001)}]{Woo2001}
{Woo}, J.-H., \& {Demarque}, P. 2001, \aj, 122, 1602

\bibitem[{{Woo} {et~al.}(2003){Woo}, {Gallart}, {Demarque}, {Yi}, \&
  {Zoccali}}]{Woo2003}
{Woo}, J.-H., {Gallart}, C., {Demarque}, P., {Yi}, S., \& {Zoccali}, M. 2003,
  \aj, 125, 754

\bibitem[{{Yi} {et~al.}(2004){Yi}, {Demarque}, \& {Kim}}]{Yi2004}
{Yi}, S.~K., {Demarque}, P., \& {Kim}, Y.-C. 2004, \apss, 291, 261

\bibitem[{{Zahn}(1991)}]{Zahn1991}
{Zahn}, J.~P. 1991, \aap, 252, 179

\end{thebibliography}

\end{document}